\RequirePackage{amsthm}

\documentclass[sn-mathphys-num, iicol]{sn-jnl}

\usepackage{graphicx}%
\usepackage{multirow}%
\usepackage{amsmath,amssymb,amsfonts,mathtools}%
\usepackage{mathrsfs}%
\usepackage[title]{appendix}%
\usepackage[table]{xcolor}%
\usepackage{textcomp}%
\usepackage{manyfoot}%
\usepackage{booktabs}%
\usepackage{algorithm}%
\usepackage{algorithmicx}%
\usepackage{algpseudocode}%
\usepackage{listings}%
\usepackage[acronym]{glossaries}
\usepackage{algpseudocode}
\usepackage{bm}
\usepackage{psfrag}
\usepackage{caption}
\usepackage{subcaption}
\usepackage{tcolorbox}
\usepackage{color} 
\usepackage{tabularx}

\definecolor{greycolor}{cmyk}{0,0,0,.8}

\theoremstyle{thmstyleone}%
\theoremstyle{thmstyletwo}%
\newtheorem{remark}{Remark}%

\theoremstyle{thmstylethree}%

\makenoidxglossaries
\newacronym{tom}{ToM}{Theory of Mind}
\newacronym{bdi}{BDI}{Belief-Desire-Intention}
\newacronym{cmm}{CMM}{Common Model of Mind}
\newacronym{tpb}{TPB}{Theory of Planned Behavior}
\newacronym{hsri}{HSRI}{Human-Social-Robot Interaction}
\newacronym{sar}{SAR}{Socially assistive robot}
\newacronym{sr}{SR}{Social robot}

\newacronym{mbc}{MBC}{Model Based Controller}
\newacronym{ml}{ML}{Machine Learning}
\newacronym{rl}{RL}{Reinforcement Learning}
\newacronym{nn}{NN}{Neural Network}
\newacronym{flc}{FLC}{Fuzzy Logic Control}
\newacronym{fcm}{FCM}{Fuzzy Cognitive Map}
\newacronym{mdp}{MDP}{Markov Decision Process}
\newacronym{pomdp}{POMDP}{Partially Observable Markov Decision Process}

\newglossaryentry{parameters}
{%
  name={$\pmb{\theta}$},
  description={Parameters for personalizing the model},
  symbol={\theta}
}

\newglossaryentry{optimal-parameter}
{%
  name={${\theta}^*$},
  description={Identified parameter},
  symbol={{\theta}^*}
}

\newglossaryentry{parameters-perception}
{%
  name={$\bm{\theta}^{\text{per}}$},
  description={Identifiable paramerters of the perception module},
  symbol={\bm{\theta}^{\text{per}}}
}

\newglossaryentry{parameters-cognition}
{%
  name={$\bm{\theta}^{\text{cog}}$},
  description={Identifiable paramerters of the cognition module},
  symbol={\bm{\theta}^{\text{cog}}}
}

\newglossaryentry{params-per-run-r}
{%
  name={$\bm{\theta}^\text{per}_r$},
  description={Values of the parameters of the perception module in run $r$},
  symbol={\bm{\theta}^\text{per}_r}
}

\newglossaryentry{params-cog-run-r}
{%
  name={$\bm{\theta}^\text{cog}_r$},
  description={Values of the parameters of the cognition module in run $r$},
  symbol={\bm{\theta}^\text{cog}_r}
}

\newglossaryentry{params-per-best}
{%
  name={$\bm{\theta}^\text{per}_{r^*}$},
  description={Values of the parameters of the perception module in run $r$},
  symbol={\bm{\theta}^\text{per}_{r^*}}
}

\newglossaryentry{params-cog-best}
{%
  name={$\bm{\theta}^\text{cog}_{r^*}$},
  description={Values of the parameters of the cognition module in run $r$},
  symbol={\bm{\theta}^\text{cog}_{r^*}}
}

\newglossaryentry{state-variable}
{%
  name={$x$},
  description={state variable},
  symbol={x}
}

\newglossaryentry{variable-domain}
{%
  name={$\mathbb{X}$},
  description={Domain of variable},
  symbol={\mathbb{X}}
}

\newglossaryentry{fuzzy-set}
{%
  name={$\tilde{X}$},
  description={Fuzzy set},
  symbol={\tilde{X}}
}

\newcommand{\dm}{decision-making}
\newcommand{\alternativeController}{rule-based controller}

\newcommand{\rld}{real-life data}
\newcommand{\pd}{perceived data}
\newcommand{\rpk}{rationally perceived knowledge}
\newcommand{\pk}{perceived knowledge}
\newcommand{\kp}{k^{\text{p}}}
\newcommand\card[1]{\left\lvert#1\right\rvert}
\newcommand{\referenceDataSetExperiment}{\cite{experiment_data_set}}

\newcommand{\belief}[1]{x^{\text{b}}_{#1}}
\newcommand{\goal}[1]{x^{\text{g}}_{#1}}
\newcommand{\emotion}[1]{x^{\text{e}}_{#1}}
\newcommand{\bias}{x^{\text{bias}}}
\newcommand{\perceivedKnowledge}[1]{x^{\text{PK}}_{#1}}

\newcommand{\realLifeData}[1]{u_{#1}}
\newcommand{\perceivedData}[1]{y^{\text{PA}}_{#1}}
\newcommand{\rationallyPerceivedKnowledge}[1]{y^{\text{RR}}_{#1}}
\newcommand{\intention}[1]{y^\text{RIS}_{#1}}
\newcommand{\action}[1]{y^\text{RAS}_{#1}}

\begin{document}

\title[Article Title]{Leveraging Systems and Control Theory for Social Robotics: A Model-Based Behavioral Control Approach to Human-Robot Interaction}


\author*[1]{\fnm{Maria} \sur{Mor\~{a}o Patr\'{i}cio}}\email{m.l.moraopatricio@tudelft.nl}

\author[1]{\fnm{Anahita} \sur{Jamshidnejad}}\email{a.jamshidnejad@tudelft.nl}

\affil[1]{\orgdiv{Control and Operations Department}, \orgname{Delft University of Technology}, \orgaddress{\street{Kluyverweg 1}, \city{Delft}, \postcode{2629 HS}, \country{The Netherlands}}}


\abstract{
Social robots (\glspl{sr}) should autonomously interact with humans, while exhibiting proper social behaviors associated to their role. 
By contributing to health-care, education, and companionship, \glspl{sr} will enhance life quality.   
However, personalization and sustaining user engagement remain a challenge for \glspl{sr},   
due to their limited understanding of human mental states.  
Accordingly, we leverage a recently introduced mathematical dynamic model of human perception, cognition, and decision-making 
for \glspl{sr}. 
Identifying the parameters of this model and deploying it in behavioral steering system of  \glspl{sr} allows to effectively personalize the 
responses of \glspl{sr} to evolving mental states of their users, enhancing long-term engagement and personalization. 
Our approach uniquely enables autonomous adaptability of \glspl{sr} by modeling the dynamics of 
invisible mental states, significantly contributing to the transparency and awareness of \glspl{sr}.
We validated our model-based control system in experiments with $10$ participants 
who interacted with a Nao robot over three chess puzzle sessions, $45-90$ minutes each. 
The identified model achieved a mean squared error (MSE) of $0.067$ (i.e., $1.675\%$ of the maximum possible MSE) in tracking beliefs, goals, and emotions of participants. 
Compared to a model-free controller that did not track mental states of participants, 
our approach increased engagement by $16\%$ on average. 
Post-interaction feedback of participants (provided via dedicated questionnaires) 
further confirmed the perceived engagement and awareness of the model-driven robot.  
These results highlight the unique potential of model-based approaches and control theory in 
advancing human-\gls{sr} interactions. 
}




\maketitle

\section{Introduction}\label{sec:intro}

\glsresetall
Over the past decades, social robotics has shown an increasing potential to improve the quality of life of humans: 
\glspl{sr} may be used in public environments, e.g., in shopping centers \cite{SR_in_mall_1, SR_in_mall_2}, museums \cite{SR_in_museum_1, SR_in_mall_2}, and airports \cite{Spencer_robot_schiphol}, to provide guidance and information to users. 
\glspl{sr} can also be deployed to assist their users, e.g., the elderly people \cite{Luperto2024, Khosla2019}, people undergoing rehabilitation \cite{HunLee2023, Tapus2008}, or children with Autism \cite{Clabaough2019, Scassellati2018ImprovingRobot, Ascencao2022}. 

In assistive contexts, in order to be effective and beneficial, \glspl{sr} must interact with their users for a prolonged period of time \cite{HunLee2023, Leite2013, Mataric2016, Khosla2019, Luperto2024}. However, most \glspl{sr} struggle to keep their users engaged in the long term. 
This prompts users to abandon using the robots over time
\cite{Khosla2019, Clabaough2019}. 
To mitigate this, \glspl{sr} often display human-like behaviors, e.g., empathy \cite{Tapus2008}, eye contact \cite{Scassellati2018ImprovingRobot}, and creativity \cite{Dellanna_1}, in order to  engage their users. 
Although users may initially be entertained by such unknown performances, i.e., they are under the novelty effect, they gradually lose interest when the 
behaviors are frequently repeated \cite{Khosla2019, Leite2013}.%

Most state-of-the-art methods steer the behavior of \glspl{sr} employing black-box methods, e.g., machine learning \cite{Bagheri2021, HunLee2023, Clabaough2019}. 
Although the behavior of \glspl{sr} has been personalized to each user (based on their fixed characteristics) in \cite{Clabaough2019, Tapus2008, Scassellati2018ImprovingRobot, Ascencao2022}, it is not adapted to their 
constantly varying mental states \cite{Mataric2016, Bagheri2021}. 
This makes the interactions to be perceived as rudimentary and artificial by the users \cite{Tapus2007a} and  subsequently hinders engaging them in the long term \cite{Mataric2016}. 
Thus, it is essential that \glspl{sr} are aware of the mental states of the users they interact with, 
just as humans (who possess theory of mind \cite{Premack1978}) are in their social interactions  \cite{Mataric2016, Tapus2007a, Bagheri2021}. 
Accordingly, we propose providing \glspl{sr} with models of the perception, cognition, and decision-making of the users. These models 
represent the dynamics of the mental states of the users. Furthermore, we propose to control the behavior of the robots based on such models. This is expected to enhance \glspl{sr} to act more human-like in \glspl{hsri}.

\subsection{Main contributions}\label{sec:contributions}

In order to advance social robotics and to particularly improve the interactions between \glspl{sr} and humans, 
we provide the following contributions:
\begin{itemize}
    \item 
   \textbf{First deployment of systems-and-control-theoretic methods for cognitive human-robot interactions:} 
    We leverage a novel dynamic mathematical model of Theory of Mind and integrate it  
    into the behavioral control system of \glspl{sr},  
    enabling real-time personalization and adaptation based on evolving mental states of humans.
    \item 
    \textbf{Development of a model-based behavioral steering system for social robots:} 
    We introduce a model-based control system that optimizes engagement, emotional well-being, and sustainability of human-\gls{sr} interactions. Furthermore, our approach improves the transparency 
    of the decision-making of \glspl{sr} and overcomes limitations of traditional model-free controllers.
    \item 
    \textbf{Empirical validation through real-life long-term human-robot interactions:} 
    Testing with $10$ participants in extended chess puzzle sessions with a Nao robot  
    demonstrated a $16\%$ increase in engagement and high model accuracy (i.e., mean squared error of $0.067$), 
    supported by objective measurements, as well as post-interaction subjective feedback from participants. 
\end{itemize}

Section~\ref{sec:methodology} of this article describes the approaches and mathematical changes applied to Mathematical Model of Mind proposed in \cite{paper1} to make it suitable for \glspl{hsri}. A case-study where the framework is applied is given in section~\ref{sec:case-study}, and the results in section~\ref{sec:results}. Finally, section~\ref{sec:conclusion} summaries the conclusions taken in this study and the recommendations for the future. 

\subsection{Related work}\label{sec:related-work}
\glsreset{tom}  

Most state-of-the-art research on steering the behavior of \glspl{sr} relies on model-free approaches. 
For example, some research projects use machine learning approaches, including reinforcement learning \cite{Clabaough2019, Gao2018, Tapus2008, Bagheri2021} and neural networks \cite{HunLee2023, Filippini2021, NN_for_perception_1, NN_for_perception_2}, to steer the behavior of the robot, adapting it to the user. 
Nonetheless, the majority of these researches \cite{Clabaough2019, Gao2018, Tapus2008} do not model the (dynamics of the) mental states of the users, nor do they try to act in accordance with these varying mental states. 
Furthermore, using neural networks in social robotics is mostly exclusive to modeling the perception of \glspl{sr} \cite{HunLee2023, Filippini2021, NN_for_perception_1, NN_for_perception_2}.
Other researches have employed rule-based approaches to determine the behavior of the \glspl{sr} \cite{Scassellati2018ImprovingRobot, Kidd2008, Schrum2019, Rossi2020}, whereas the input or the evaluation of these rules 
is based on metrics that mainly reflect variables defined for the interactions 
(e.g., task performance \cite{Scassellati2018ImprovingRobot, HunLee2023} and user engagement \cite{Kidd2008}), rather than on the mental states of the users and the dynamics of these states. 

As for model-based approaches, some works have recently developed biological models of the brain \cite{Patacchiola2020, Zeng2020}, where the behavior of the \gls{sr} is generated by a brain-inspired architecture. 
This is mainly proposed for \glspl{sr} to learn trust \cite{Patacchiola2020} or to distinguish false beliefs \cite{Zeng2020}. Nevertheless, these models create the behavior of the \gls{sr}, rather than 
representing the user cognition and enabling the \gls{sr} to be aware of this cognition. 

Scassellati in \cite{Scassellati2002} poses that humanoid robots should manifest a \gls{tom}, which is the capability of rational agents to be aware of and to reason about the perception, beliefs, desires, and decision-making of other rational agents \cite{Scassellati2002, Premack1978}. 
Since then, some researches have applied the concept of \gls{tom} for \glspl{sr} \cite{Zeng2020, Dissing2020, Vossen2018, Patacchiola2020, Lee2019}. However, they only focused on one particular aspect of \gls{tom}, 
i.e., on modeling the belief (which relates to the perception, rather than on cognition and decision-making processes) of the users. 
Recently, a steering system for \glspl{sr} that consists in a white-box cognitive model based on the 
BDI (Belief-Desire-Intention) framework \cite{Bratman1987IntentionReason} has been proposed and deployed in 
\glspl{hsri} \cite{Dellanna_2}. In addition to enabling adaptive and transparent responses to the users, such a model-based steering system enhances the engagement of the users \cite{Dellanna_2}.%

In our earlier work \cite{paper1}, we have proposed a mathematical framework that 
models the perception, cognition, and decision-making of humans, following the core \gls{tom} principles. 
In this paper, we refer to this model as Mathematical Model of Mind, or briefly MMM. 
In particular, we leverage the MMM to be used by \glspl{sr} to track and act upon the mental states of their human users, by integrating the model within a control system that is suitable for the purposes of \glspl{hsri}. 

\section{Methodology}
\label{sec:methodology}
\glsresetall

In this section we describe our proposed model-based control framework for \glspl{sr}. In particular,  
we explain how the dynamic mathematical model of perception, 
cognition, and decision-making introduced in \cite{paper1} i.e., the MMM, 
is leveraged to be used in a model-based control framework for \glspl{sr}.%

In \cite{paper1}, a path diagram, which illustrates variables that are connected via 
processes and weighted linkages, 
has been proposed to provide 
a white-box representation for the inter-dynamics of the (invisible) mental states and their 
contribution to the (visible) actions of humans, based on the \gls{tom}. 
The variables represented in the path diagram can be either static variables (i.e., variables without a memory 
or impact from their previous realizations) and dynamic variables (i.e., variables with a memory 
that are impacted by the past realizations). 
The static variables of the path diagram are real-life data, perceived data, rationally perceived knowledge, intention, and action, and the dynamic variables include beliefs, goals, emotions, perceived knowledge, and biases.%

The path diagram is then formulated in \cite{paper1} as a mathematical model, which we refer to as MMM, 
using systems theory. MMM includes three modules: 
perception, which models how real-life data is perceived by humans to impact their mental states; 
cognition, which represents the dynamic evolution of the mental states of humans; 
and \dm, which determines the actions of the humans based on their mental states.%

The dynamic processes of MMM have been developed in the state space framework, 
thus are represented in terms of a set of inputs, outputs, 
state variables (i.e., dynamic  variables that represent the internal mental 
states at any given time and that evolve according to a differential or difference equation), 
and dynamic auxiliary variables (i.e., dynamic  variables that are additionally included in the model to allow for a 
realistic and precise bridging of the different modules and state variables). 
All the dynamic variables (i.e., beliefs, goals, and emotions, which are the 
state variables, as well as the bias and the perceived knowledge, which are the dynamic auxiliary variables) are 
updated within the cognition module.%

The three modules of perception, cognition, and \dm\ are sequentially integrated, i.e., 
the output variable of the perception module is injected as input into the cognition module, 
which updates all the dynamic variables of the model. 
Then, the updated state variables beliefs and goals are injected into the \dm\ module, which outputs the action. 
Note that although the state variable emotion does not directly impact the \dm\ processes, 
it does impact the updated state variables beliefs and goals via their inter-dynamics.%

The state space representation of MMM allows the model to be embedded within control-theoretic approaches, allowing to steer the \glspl{hsri} transparently with guarantees on the performance and constraints satisfaction. 
Next, we describe how the MMM is leveraged for incorporation in a control-theoretic method for steering the behavior of \glspl{sr}. The model is represented in a discrete time framework.%


\subsection{Leveraging MMM for model-based control of SRs}
\label{sec:MMM_leverage}
We introduce improvements to the formulation of MMM \cite{paper1} that make 
the resulting model suited for being adopted by control-theoretic methods in real-life implementations of \gls{hsri}.%

Note that according to the neuroscience findings although 
perception, cognition, and \dm\ are interconnected, their processes occur with different speeds \cite{frequencies_of_perception_and_motor}. 
More specifically, the perception processes (i.e., the transition of the external stimuli into a perception) occur almost three times faster than the cognition and motor processes (i.e., the transition of the decisions into actions) \cite{sience_frequencies_of_brain_waves}. 
Moreover, the perception depends on the external stimuli and should be updated every time a new external stimulus is received. 
However, the cognition evolves not only due to the perception but also under the impact of the inter-dynamics of the mental states. 
Therefore, the cognition may be updated with a frequency different from the frequency of capturing the external stimuli. 
Consequently, in our next discussions, we have used  different discrete time steps for the perception (where the discrete time 
steps are indicated by $\kp$) and for the cognition and \dm\ (where $k$ is used to show the corresponding 
discrete time steps).%

The improvements made in different modules of MMM and the motivation for introducing them are explained next.%

\label{sec:methods-modeling}

\bmhead{Perception module}

\begin{figure}
    \centering
    \includegraphics[width=0.48\textwidth]{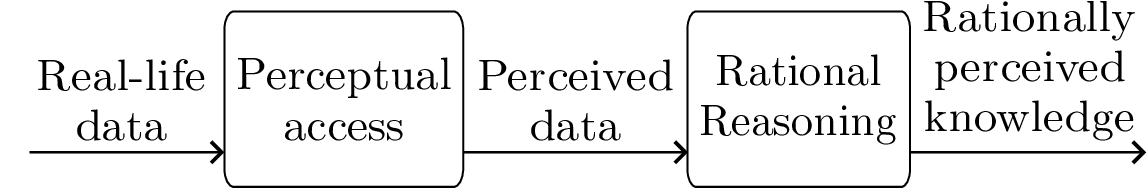}
    \caption{Perception module.}
    \label{fig:perception-module} 
\end{figure}

Perception is the process of translating \rld\ into knowledge of the situation. 
Perception in MMM is mathematically modeled with three variables, i.e., \rld, \pd, and \rpk, and through two linked sub-processes, perceptual access  and rational reasoning. 
The perceptual access is the transformation of the \rld\ into perceived data, which is the data captured by the human. The rational reasoning is the transition of the perceived data into \rpk, which is knowledge that would ideally be reasoned by that individual when not biased by current mental states.
Figure~\ref{fig:perception-module} illustrates these sub-processes and their input-output pairs.%

In order to leverage MMM for model-based control of \glspl{sr} when used in the context of \gls{hsri}, we propose the following 
improvements (highlighted in the boxes and elaborated in the text) for the perception module of MMM.

\begin{tcolorbox}[colback=gray!5!white,
                  colframe=gray!75!black,
                  boxrule=1pt, 
                  arc=0.6em, 
                  boxsep=.5mm, 
                  title = Specifying the perceptual access mathematically]
In \cite{paper1}, the perceptual access sub-process is
represented via a generic function $f^{\text{PA}}(\cdot)$ without specification regarding its class or characterization.
We fill this gap to enable deployment on SRs by representing the perceptual access as a filter where the output (i.e., perceived data) equals the input (i.e., \rld) when the input is captured, and, otherwise, equals the last captured value.
Our solution presents a transparent way to mathematically represent the perceptual access, enhancing the white-box nature of MMM, and facilitating its usage by \glspl{sr}. 
\end{tcolorbox}  

Suppose that $\mathcal{R}$ is an ordered set which includes all the \rld\ in a given \gls{hsri} context. 
At the discrete time step $\kp$, the value of the elements of $\mathcal{R}$ are denoted by $\realLifeData{i} (\kp)$ with  $i = 1, \ldots, \card{\mathcal{R}}$. 
Moreover, $\perceivedData{i} (\kp)$, with $i = 1,\ldots, \card{\mathcal{R}}$, 
is the $i^{\text{th}}$ element of the output of the perceptual access (i.e., 
$i^{\text{th}}$ element of the perceived data) at time step $\kp$, which is determined based on $\realLifeData{i} (\kp)$. Whenever the human captures the $i^{\text{th}}$ element of \rld, then $\perceivedData{i} (\kp)$ is replaced by the value of this element, $\realLifeData{i} (\kp)$. Otherwise, the value of $\perceivedData{i} (\kp)$ keeps its most recent value. The mathematical representation of the perceptual access is:
\begin{align}
    \perceivedData{i} (\kp) = 
    \left\{
    \begin{array}{ll}
      \realLifeData{i} (\kp),       &  \text{if $\realLifeData{i} (\kp)$ was captured}\\
      \perceivedData{i} (\kp - 1),  &  \text{otherwise}
    \end{array}
    \right.
\label{eq:perception-general-pa}
\end{align}

\begin{tcolorbox}[colback=gray!5!white,
                  colframe=gray!75!black,
                  boxrule=1pt, 
                  arc=0.6em, 
                  boxsep=.5mm, 
                  title = Decoupling \& parameterizing the function that models the rational reasoning]
In \cite{paper1}, the rational reasoning sub-process is represented via a generic function $f^{\text{RR}}(\cdot)$, 
which remains abstract, without specifying its class or characterization. 
We address this gap for deployment on SRs, which should understand how humans reason about perceived data, by decoupling and parameterizing the function that maps the input (i.e., the perceived data) of the sub-process to its output (i.e., rationally perceived knowledge). 
Our solution further enhances the white-box nature of MMM, thus its transparency and traceability, 
and yields a practical and computationally efficient 
way for personalizing the perception process to users. 
\end{tcolorbox}  

The perceived data is injected as input into the rational reasoning sub-process 
to generate as output the rationally perceived knowledge (see Figure~\ref{fig:perception-module}). 
Thus, the outputs $\perceivedData{i} (\kp)$ of the perceptual access 
are the inputs to the rational reasoning. 
The output of the rational reasoning, i.e., rationally perceived knowledge, is composed of elements 
$y_j^{\text{RR}}(\kp)$, where $j = 1, \ldots, \card{\mathcal{K}}$ 
and $\mathcal{K}$ is the ordered set that embeds all potential realizations of 
the rationally perceived knowledge in the given \gls{hsri} context.%

We consider \glspl{hsri} that require selective attention from humans 
(e.g., a particular educational or entertaining joint task by the \gls{sr} and the human). 
In other words, the human focuses on a specific stimulus or input at a time. 
This allows us to decouple the input-output mapping (i.e., the mapping that shows the 
impact of input $\perceivedData{i} (\kp)$ on output $\rationallyPerceivedKnowledge{j} (\kp)$) 
in the mathematical modeling of the rational reasoning. 
Considering the additivity condition (which allows for simplicity and efficiency of the computations) 
on this input-output mapping, we have:
\begin{align}
     y_j^{\text{RR}}(\kp) &= \sum_{i = 1}^{\lvert\mathcal{P}\rvert} f^{\text{RR}}_{ij} (\perceivedData{i} (\kp); \bm{\theta}_{ij}^{\text{RR}}), \quad \forall j \in \mathcal{K} 
\label{eq:perception-general-rr}
\end{align}
In \eqref{eq:perception-general-rr}, the sub-process is modeled as the summation of various 
parameterized functions, each mapping one input of the sub-process to its corresponding output. 
More specifically, the contribution of the 
$i^{\text{th}}$ element $\perceivedData{i} (\kp)$ of $\mathcal{P}$ through the 
rational reasoning process into the $j^{\text{th}}$ element $y_j^{\text{RR}}(\kp)$ of $\mathcal{K}$ 
(i.e., the $j^{\text{th}}$ element of the rationally perceived knowledge at time step $\kp$) 
is modeled by the function $f^{\text{RR}}_{ij}: \mathcal{P} \rightarrow \mathcal{K}$. 
Note that the function $f^{\text{RR}}_{ij}(\cdot)$ is defined as parameterized, i.e., $\bm{\theta}_{ij}^{\text{RR}}$ is a parameter vector that 
should be identified per user for all indices $i$ and $j$. 
This allows for more flexibility in the mathematical model and for the ease of personalization per user.%

\begin{remark}
Sets $\mathcal{P}$, and $\mathcal{K}$  
are defined as ordered sets, because the inputs and outputs in \eqref{eq:perception-general-pa} and \eqref{eq:perception-general-rr} 
need to be distinguishable to be associated to the corresponding parameterized function of the model.   
\end{remark}

\begin{figure}
    \centering
    \fontsize{8}{10}
    \includegraphics[width=0.48\textwidth]{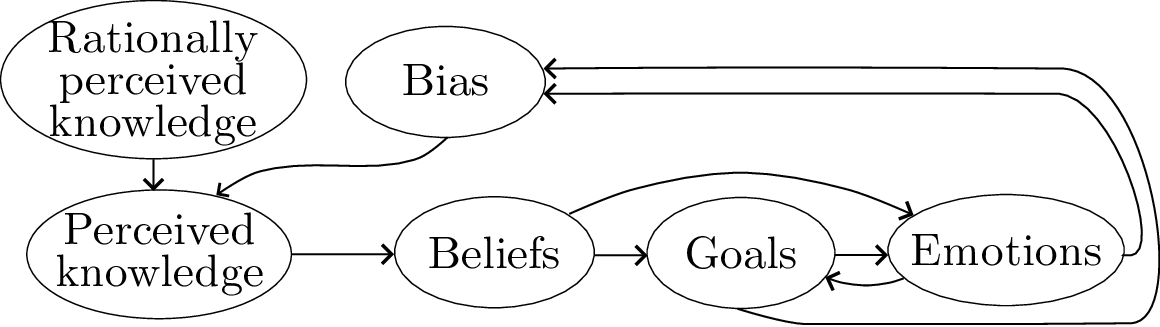}
    \caption{Cognition module.}
    \label{fig:cognition-module} 
\end{figure}

\bmhead{Cognition module}
Cognition refers to the processes that use the output of the perception process, 
i.e., the rationally perceived knowledge, to update the mental states of the human 
that will determine the decision-making and action-planning of that human. 
In \cite{paper1}, using a state space framework, the dynamic state and auxiliary variables (shown by $\glssymbol{state-variable}(k)$ and distinguished by subscript indices) of the cognitive module 
are updated at each discrete time step $k+1$, based on their values at the previous discrete time step $k$. 
Note that different indices do not only refer to different dynamic variables 
(e.g., belief versus emotion) but also to different elements for one variable (e.g., different beliefs that 
play a role in the given \gls{hsri} context). 
We have\footnote{Note that indices used in different sections of the paper (e.g., $i$ and $j$) should be treated 
as local variables, i.e., there is no relevance between these indices used, for instance, in the mathematical discussions for the cognition module and the same indices used elsewhere in the mathematical discussions for the perception or decision-making modules.}:

  \begin{align}
    \label{eq:update-FCM}
        \glssymbol{state-variable}_i(k+1)
         = 
        f_i\left(\glssymbol{state-variable}_i(k) \right) &+ 
        \sum_{i \neq j} 
        w_{ji} (k) \glssymbol{state-variable}_j(k)
        \nonumber
        \\
        & +
       \sum_{\ell \in \mathcal{K}} z_{\ell i}(k) y_\ell^{\text{RR}}(k)
  \end{align}
%
where the first term on the right-hand side of 
\eqref{eq:update-FCM} models the impact of the previous realization of the variable, 
i.e., of $\glssymbol{state-variable}_i(k)$, on its updated realization, i.e., 
on $\glssymbol{state-variable}_i(k + 1)$.
The second term on the right-hand side of \eqref{eq:update-FCM} represents the inter-dynamics of the state and dynamic auxiliary variables, with 
$w_{ji} (k)$ 
the relative weight showing the impact of variable $\glssymbol{state-variable}_j(k)$ (called the influencing variable) 
on updated variable $\glssymbol{state-variable}_i(k+1)$ (called the influenced variable) at discrete time step $k$. The third term on the right-hand side of \eqref{eq:update-FCM} represents the influence of 
those inputs that are external to the cognitive module, i.e., the influence of the output of the perception module, the rationally perceived knowledge, on the dynamic variable perceived knowledge (see Figures~\ref{fig:perception-module} and \ref{fig:cognition-module}). 
\begin{remark}
 Note that \eqref{eq:perception-general-rr} estimates $y_\ell^{\text{RR}}(\kp)$ more frequently than the update frequency of the cognition module. Thus, in \eqref{eq:update-FCM} $y_\ell^{\text{RR}}(k)$ is the most recent value of $y_\ell^{\text{RR}}(\kp)$ that has been estimated prior to time step $k$. 
\end{remark}

The weights $w_{ji} (k)$ and $z_{\ell i} (k)$ in \eqref{eq:update-FCM} are, in general, functions of both the influencing and influenced variables at the current time step $k$, i.e.:
\begin{subequations}
\begin{align}
\label{eq:FCM-weights}
&w_{ji} (k) 
 = f_{ji}^\text{W} \left( \glssymbol{state-variable}_i(k) , \glssymbol{state-variable}_j(k)\right)
\\
&z_{\ell i} (k) = 
\\
& 
\left\{
\begin{array}{ll}
  f_{\ell i}^\text{Z} \left( \glssymbol{state-variable}_i(k) ,  y_\ell^{\text{RR}}(k)\right),   &  \text{$\glssymbol{state-variable}_i(k)$ is perceived knowledge}\\
  0,   &  \text{otherwise}
\end{array}
\right.
\nonumber
\end{align}
\end{subequations}
with $f_{ji}^\text{W}(\cdot)$ and $f_{\ell i}^\text{Z}(\cdot)$ functions that estimate the weights based on the values of the influencing and influenced variables per time step.%

We propose the following improvements (highlighted in the boxes and elaborated in the text) 
to leverage the cognition module of MMM for real-life applications in \gls{hsri}.%

\begin{tcolorbox}[colback=gray!5!white,
                  colframe=gray!75!black,
                  boxrule=1pt, 
                  arc=0.6em, 
                  boxsep=.5mm, 
                  title = Incorporating the different update frequencies of dynamic state or auxiliary variables]
Not all dynamic variables in MMM are in general updated with the same frequency. 
This difference should thus be incorporated in the update equation \eqref{eq:update-FCM}. 
Accordingly, we propose to represent the first term, i.e., $f_i\left(\glssymbol{state-variable}_i(k) \right)$,  
 as a weight $w_{ii}$ multiplied by the most recent realization of the variable, $\glssymbol{state-variable}_i(k)$. 
While each weight $w_{ii}$ models how fast a variable responds to its last realization, the relative values of all the 
weights in \eqref{eq:update-FCM} incorporate the difference of the update frequencies for different variables.%
\end{tcolorbox}  

In MMM \cite{paper1}, the representation of function $f_i(\cdot)$, as well as 
the relative update frequencies of the dynamic variables 
were not discussed. 
As the processes that result in the formation of beliefs\footnote{Beliefs correspond to the 
internal representations of stimuli, arising from the perception of a rational agent, 
and thus vary in time as the perception does. Knowledge (e.g., moral values, political beliefs, etc.) 
that is fixed or that varies very slowly is deemed pieces of general world knowledge of the agent).}
are included in the perception module,  
and the perception processes are faster than the cognition processes \cite{sience_frequencies_of_brain_waves}, the beliefs are in general updated more frequently than the emotions and goals. 
Accordingly, we propose:
\begin{align}
\label{eq:update-FCM-self-influence}
    \begin{split}
    f_i(\glssymbol{state-variable}_i(k)) =
    w_{ii}  \glssymbol{state-variable}_i(k)
    \end{split}
\end{align}
By adjusting the value of $w_{ii}$, one models how intensely the variable $\glssymbol{state-variable}_i(k)$ changes per discrete time step (i.e., how quickly the variable responds to its previous realization). 
Moreover, the relative value of $w_{ii}$ and $w_{ji} (k)$ in \eqref{eq:update-FCM} 
incorporates  the difference in the frequency of the self-impact 
and the impact by other variables. 

For the beliefs, in particular, we use $w_{ii}=0$ since the beliefs per time step 
depend on the current stimuli, and are generated by the perceived knowledge (see \autoref{fig:cognition-module}). Any internal impacts 
that may affect the beliefs (e.g., the impact of the most recent goals and emotions) 
are incorporated through the impact of the auxiliary variable bias on the perceived knowledge 
(see \autoref{fig:cognition-module})

If the weight $w_{ii}$ tends to $1$, it leads to instability of \eqref{eq:update-FCM} \cite{discrete_systems_reference}. 
To ensure the stability of the model, all weights $w_{ii}$ and $w_{ji}$ should be chosen as to assure that all 
the eigenvalues of the corresponding matrix $W$ (i.e., a matrix with weights $w_{ii}$ in its diagonal element positions $(i,i)$ and weights $w_{ij}$ with $i \neq j$ in positions $(i,j)$) are within the unit circle \cite{discrete_systems_reference}. 


\begin{tcolorbox}[colback=gray!5!white,
                  colframe=gray!75!black,
                  boxrule=1pt, 
                  arc=0.6em, 
                  boxsep=.5mm, 
                  title = Modeling the inter-dynamics of mental variables via piecewise-constant functions]
In \eqref{eq:FCM-weights}, as is given in \cite{paper1}, 
$f_{ji}^{\text{W}}(\cdot)$ is presented as generic and abstract, while 
specific characteristics in formulating $f_{ji}^{\text{W}}(\cdot)$ 
for meeting desired criteria by MMM remain vague.  
We address this gap by proposing a mathematical  representation for $f_{ji}^{\text{W}}(\cdot)$ 
that provides a balanced trade-off between computational efficiency 
and adequate flexibility for illustrating the interpersonal differences of humans, 
that is always striven for in \glspl{hsri}. 
Accordingly, we propose to define $f^{\text{W}}_{ji}(\cdot)$ as a piecewise function with respect to the influenced variable $x_i(k)$ and
influencing variable $x_j(k)$, where these constant values 
should be identified per human. 
\end{tcolorbox}  

Suppose that $\mathcal{X}_i$ and $\mathcal{X}_j$ are the admissible sets for all the potential realizations of the state/auxiliary  variables $\glssymbol{state-variable}_i$ and $\glssymbol{state-variable}_j$, respectively. 
Moreover, $\overline{\mathcal{X}}_i \subseteq \mathcal{X}_i$ and $\underline{\mathcal{X}}_i \subseteq \mathcal{X}_i$, 
with $\overline{\mathcal{X}}_i \cup \underline{\mathcal{X}}_i = \mathcal{X}_i$ and 
$\overline{\mathcal{X}}_i \cap  \underline{\mathcal{X}}_i = \emptyset$, 
include two distinct ranges for the influenced variable. For instance, they 
may be interpreted as ranges of undesirable and desirable status for the variable (e.g., spectrum of 
feeling bored and spectrum of feeling interested for the emotion state variable ``boredom''). 
We also consider $\mathcal{X}_{j, 1}, \ldots, \mathcal{X}_{j, n} \subseteq \mathcal{X}_j$ with 
$\cup_{m = 1}^n \mathcal{X}_{j, m} = \mathcal{X}_j$ and 
$\mathcal{X}_{j, m} \cap \mathcal{X}_{j, o} = \emptyset $ for $ m,o = 1, \ldots, n$ and 
$m\neq o$ as distinct ranges of the influencing variable, 
where in each of these sub-sets the impact of the influencing variable on the influenced variable may be considered as constant or non-significant. 
Then for $x_j(k) \in \mathcal{X}_{j, m}$ with $m = 1, \ldots, n$, we propose the following representation for the function $f_{ji}^{\text{W}}(\cdot)$: 
\begin{align}
    \label{eq:Definition_weight}
    w_{ji}(k) = 
    \left\{
    \begin{array}{lr}
         \zeta_{j,m}, & \qquad x_i(k) \in \overline{\mathcal{X}}_i\\
         \xi_{j,m}, & \qquad x_i(k) \in   \underline{\mathcal{X}}_i  
    \end{array}
    \right.
\end{align}
where $\zeta_{j,m}$ and $\xi_{j,m}$ are fixed parameters that should be identified per user of the \gls{sr} for all values of the indices $j$ and $m$. 
Note that \eqref{eq:Definition_weight} provides a simple, easily implementable, and computationally 
efficient formulation for real-life applications on the model for \glspl{sr}, while any potential nonlinearity in the inter-dynamics of the 
influencing and influenced variables will properly be represented by proper tuning of 
the number $n$ of subsets $\mathcal{X}_{j, m}$.%

\begin{figure}
    \centering
    \includegraphics[width=0.48\textwidth]{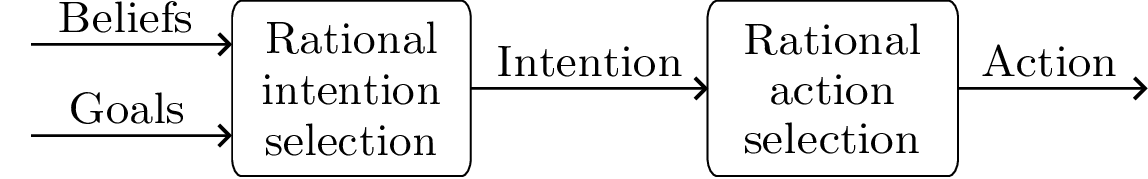}
    \caption{Decision-making module.}
    \label{fig:decision-making}
\end{figure}

\bmhead{Decision-making module}
Decision-making in humans is the process of deciding about actions, depending on 
the mental states, namely beliefs and goals. The \dm\ process in MMM has been modeled via two sub-processes that are 
linked through a static auxiliary variable, called the intention. 
These two sub-processes are the rational intention selection (i.e., the conversion of the beliefs and goals into intentions) 
and the rational action selection (i.e., the conversion of the intentions into actions).  
Figure~\ref{fig:decision-making} shows a simplified illustration of the decision-making module in MMM.

In order to be able to model the \dm\ of humans in real-life \glspl{hsri} using MMM, 
we propose the following improvements (highlighted in the box and elaborated in the text).%

\begin{tcolorbox}[colback=gray!5!white,
                  colframe=gray!75!black,
                  boxrule=1pt, 
                  arc=0.6em, 
                  boxsep=.5mm, 
                  title = Parameterized formulation of rational intention selection based explicitly on beliefs and goals]
In MMM \cite{paper1}, an explicit formulation based on the beliefs and goals for the rational intention selection sub-process 
 has not been provided. We address this gap by proposing a piecewise affine parameterized 
formulation that outputs the values for the intensity of intentions, based on the values realized for the beliefs and goals.
This formulation provides a functional way to individually identify the impact of the state variables on the intention, and to estimate their weighted summation to obtain the intentions. 
This yields an explainable, straightforward, and easily personalizable model for 
representing the rational intention selection of humans for \glspl{sr}.%
\end{tcolorbox}  
%

Suppose that $\intention{i} (k)$ is used to show the $i^{\text{th}}$ element of the intention 
at time step $k$, with $i \in \mathcal{I}$ and $\mathcal{I}$ a set including all integer indices for the intention elements. 
In fact, per time step $k$, the rational intention selection sub-process outputs vector $\bm{y}^{\text{RIS}}(k)$ 
that includes $\intention{i} (k)$ for all $i \in \mathcal{I}$.
Element $\intention{i} (k)$ quantifies the strength of intention $i$ that depends on the realized values of the beliefs and goals at time step $k$ that influence that intention. 
For each $i \in \mathcal{I}$ we have:
\begin{align}
\label{eq:dm-activation-functions}
\intention{i} (k) =
\sum_{j\in\mathcal{B}}
\theta^\text{RIS}_{ij} x_j (k)
+
\sum_{\ell\in\mathcal{G}}
\theta^\text{RIS}_{i\ell} x_\ell (k)
+ 
\theta^\text{RIS}_i 
\end{align}
where $\mathcal{B}$ and $\mathcal{G}$ are the sets of all integer indices for the beliefs and goals, respectively. 
Moreover, $\theta^\text{RIS}_{ij}$, $\theta^\text{RIS}_{i\ell}$, and $\theta^\text{RIS}_i$ are parameters that relatively weigh the influence of, respectively, one's beliefs, goals, and (fixed) neutral intention 
(i.e., the intention that would be realized when beliefs and goals are zero) on the $i^{\text{th}}$ intention.

Finally, the rational action selection sub-process 
receives $\bm{y}^\text{RIS}(k)$ as input, and outputs vector $\bm{y}^\text{RAS}(k)$, 
which includes all elements of the action taken by the human at time step $k$. 
Each element of the action vector is simply a binary variable, with  
$1$ implying that the action is selected, and $0$ implying that the action is excluded. 
The process of transitioning the quantified strength of the intentions into binary actions 
via rational action selection depends on various practical aspects of the \gls{hsri} setup, 
such as the level of abstraction of the actions (e.g., whether only one or multiple actions are required to fulfill an intention) and any constraints that restrict the actions. In general, this process is defined depending on the \gls{hsri} setup.%



\subsection{Identification of the model}
\label{sec:methods-identification}

The formulations given in Section~\ref{sec:methods-modeling} for MMM show that the model includes parameters that need to be identified (preferably personalized for each user) using approaches based on, e.g., gradient descent \cite{Bishop2024} or genetic algorithms \cite{kramer2017genetic}.%

MMM has been presented comprehensively so that it can be adopted in different contexts involving \glspl{hsri}. To achieve this, it includes multiple parameters. 
For a particular context, however, not all the modeled inter-dynamics, and thus parameters, are relevant. 
Therefore, a crucial consideration in the identification of MMM for a particular context is to avoid 
redundancy of the parameters, 
particularly because these parameters 
mostly correspond to real-life concepts related to 
the mental processes.%

This redundancy particularly occurs when the structure of the model is too complex for the size of the dataset that is available and relevant in a particular context to train the parameters. 
While performing longer and wider ranges of interactions may allow the collection of more data, this is often not desirable/possible in \glspl{hsri}. Moreover, larger datasets do not necessarily guarantee to circumvent the issue, especially when the collected data is not diverse enough for the complexity level of the model. 

Hence, based on the context and the concepts that define the realizations of the variables, linkages, and 
parameters of MMM, one may need to gradually, and systematically, simplify the model structure. 
While this is context-dependent, we propose two main steps that together allow evaluating and reducing the redundancies in the model 
every time it is simplified for a given context. 
This simplification, which includes removing some linkages, variables, and parameters, alongside these two steps can ensure proper identification of the MMM.
These steps include parameter redundancy evaluation, given in Algorithm~\ref{alg:test-params-ided}, and  
 a multi-stage identification procedure with warm start, explained via Algorithms~\ref{alg:sep-per-warm-start}--\ref{alg:sep-per-B}.%

\paragraph{Evaluating the presence of redundant parameters in MMM}

In order to evaluate whether the model needs to be further simplified,  
the parameters that are unequivocally identifiable in the current structure of MMM are established using Algorithm~\ref{alg:test-params-ided}. 
The MMM will be simplified 
by gradual, systematic removal of variables and linkages whose corresponding parameters have not been identified unequivocally. This continues until all remaining parameters are unequivocally identifiable.%

Note that the identification procedure involves solving generally non-convex optimization problems (i.e., minimizing the error of the values that are generated by the model with regard to the corresponding values collected from real-life \glspl{hsri}). Therefore, the values that are returned by the optimizer may correspond to local, 
rather than global, optima. 
To circumvent the issue properly, 
the identification optimization is performed $n^\text{run}$ independent times, each time with different starting values for the parameters.
Then, the $n^\text{opt}$ (where $n^\text{opt} \leq n^\text{run}$) sets of parameters that yielded the least realized value for the error function in the validation stage are chosen. 
In case the standard deviation of the $n^\text{opt}$ chosen values per parameter 
is small enough (i.e., less than a small threshold), the parameter is deemed identified.%

\begin{algorithm}
\caption{Procedure to determine the parameters of MMM that are unequivocally identified.} 
\label{alg:test-params-ided}
\begin{algorithmic}[1]
\Require
\Statex $r$: Index of current identification run
\Statex $p$: Index of parameter
\Statex $n^\text{run}$: Number of independent identification runs
\Statex $n^\text{opt}$: Number of optimal identification runs considered in assessment
\Statex $n^{\text{par}}$: Number of parameters considered
\Statex $\theta_{p, r}$: Identified value of parameter of index $p$ in run $r$
\Statex $\bm{\theta}_p^*$: Vector of dimension $n^\text{opt}$ including the identified values of parameter of index $p$ in the $n^\text{opt}$ runs with minimum optimization cost 
\Ensure
\State Run identification procedure $n^\text{run}$ times, each time with different, random starting values for all $n^{\text{par}}$ parameters
\State $\bm{r}^* \gets$ indices of $n^\text{opt}$ runs with minimum optimization cost in the validation stage
\For{$p \in\{1, ..., n^{\text{par}} \}$}
    \State $\bm{\theta}_p^* \gets \theta_{p, r} \qquad \forall r \in \bm{r}^*$ 
    \State $\sigma_p \gets$ standard deviation of $\bm{\theta}_p^*$
    \If{$\sigma_p \leq $ threshold}
        \State $\theta_p$ was unequivocally identified
        \Else
        \State $\theta_p$ was not unequivocally identified
        \EndIf
\EndFor
\end{algorithmic}
\end{algorithm}



\paragraph{Multi-stage optimization with warm start}
After collecting measurements for the beliefs, goals, emotions, and actions of a person, for the sake of simplicity, the identification optimization may be decoupled for the perception, cognition, and \dm\space modules. 
By knowing the beliefs, goals, and actions (i.e., the inputs and outputs of the \dm\ module, as it is illustrated in \autoref{fig:decision-making}), it is straightforward to decouple this module from the other two modules. 
Contrarily, the perception and cognition modules are less trivial to decouple since the output of the perception module that is then injected into the cognition module is an auxiliary variable, and thus cannot be measured, e.g., by asking people about it in the course of \glspl{hsri}.%

\begin{figure}
    \centering
    \includegraphics[width=0.47\textwidth]{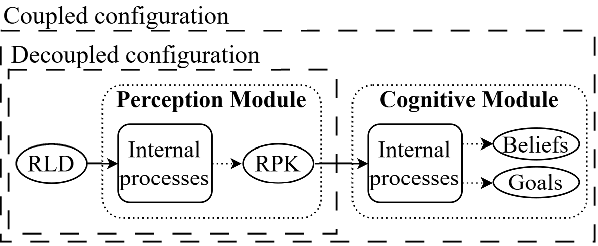}
    \caption{The two configurations used in the two-stage identification of the \gls{tom} model (RLD stands for real-life data and RPK for rationally perceived knowledge). The \textit{coupled configuration} includes both the perception and cognition modules, while the \textit{decoupled configuration} includes only the perception module.}
    \label{fig:sep-per} 
\end{figure}

Therefore, we propose dividing the identification procedure into multiple optimization stages.
To achieve this, we introduce two approaches, A and B, to perform the identification optimization. These approaches are given in Algorithms~\ref{alg:sep-per-A} and \ref{alg:sep-per-B}. 
Both approaches start with a preliminary identification of the parameters of the perception module, given by Algorithm~\ref{alg:sep-per-warm-start}, whose outcome is then used as a warm start. 
In this case, the perception module of the model is decoupled from the cognition module and is solely considered 
(called the \textit{decoupled configuration} in \autoref{fig:sep-per}).  
In the \textit{decoupled configuration}, the inputs and outputs are, respectively, real-life data and rationally perceived knowledge. 
Since the values of the rationally perceived knowledge are not measurable, the pre-identification for 
the \textit{decoupled configuration} is performed under the assumption that each element of the rationally perceived knowledge for time step $k$ is the same as the corresponding element of the belief for the same time step $k$ (i.e., assuming the bias is null).%

\begin{algorithm}
\caption{Generating the warm start for the parameters of the perception module}\label{alg:sep-per-warm-start}
\begin{algorithmic}[1]
\Require
\Statex $r$: Index of current identification run
\Statex $r^*$: Index of run with minimum optimization cost
\Statex $n^\text{run}$: Number of independent runs
\Statex $\glssymbol{params-per-run-r}$: Vector containing identified values for parameters of perception module in run $r$
\Statex $\bm{\theta}^{\text{per-ws}}:$ Vector containing warm start of parameters of perception module
\Statex $y_\ell^{\text{RR}}(k)$: Realized value of element $\ell$ of rational reasoning at time step k
\Statex $x_\ell(k)$: Realized value of element $\ell$ of belief at time step k
\Statex $\mathcal{B}$: Set of all integer indices of beliefs. 
\Ensure
\State Assume $y_\ell^{\text{RR}}(k)$ = $x_\ell(k) \qquad  \forall \ell \in \mathcal{B}$
\For{$r \in \{1, ..., n^\text{run}$\}}
    \State Randomly initialize $\glssymbol{params-per-run-r}$
    \State Optimize $\glssymbol{params-per-run-r}$  within \textit{decoupled configuration}
\EndFor
\State $\bm{\theta}^{\text{per-ws}} \gets \glssymbol{params-per-best}$ for run $r^*$ with minimum optimization cost
\end{algorithmic}
\end{algorithm}

After obtaining a warm start for the parameters of the perception module, one of the following two approaches is adopted to identify the entire model.%

\bmhead{Approach A}

The perception and cognition modules are simultaneously identified, considering the \textit{coupled configuration} 
shown in \autoref{fig:sep-per}. 
The starting values in the identification optimization for the parameters of the perception module  
are those obtained in the pre-identification stage explained above. 
The starting values for the parameters of the cognition module are randomly selected. 
Approach A is described in detail in Algorithm~\ref{alg:sep-per-A}.  
\begin{algorithm}
\caption{Model identification with warm-start - Approach A}\label{alg:sep-per-A}
\begin{algorithmic}[1]
\Require
\Statex $r$: Index of current identification run
\Statex $n^\text{run}$: Number of identification runs 
\Statex $\bm{\theta}^{\text{per-ws}}$: Vector containing warm start of parameters of perception module obtained in Algorithm~\ref{alg:sep-per-warm-start}
\Statex $\glssymbol{params-per-run-r}$
: Vector containing identified values for parameters of perception module in run $r$
\Statex $\glssymbol{params-cog-run-r}$: Vector containing identified values for parameters of cognition module in run $r$
\Ensure
\For{$r \in \{1, ..., n^\text{run}$\}}
\State $\glssymbol{params-per-run-r} \gets \bm{\theta}^{\text{per-ws}}$ (initialization)
\State Randomly initialize $\glssymbol{params-cog-run-r}$. 
\State Optimize $\glssymbol{params-per-run-r}$ and $\glssymbol{params-cog-run-r}$ within \textit{coupled configuration}
\EndFor
\State $\glssymbol{params-per-best}, {\glssymbol{params-cog-best}} \gets \glssymbol{params-per-run-r}, \glssymbol{params-cog-run-r}$ for run $r$ with minimum optimization cost
\end{algorithmic}
\end{algorithm}

\bmhead{Approach B}

Before identifying all parameters of the perception and cognition modules together, 
a pre-identification for the parameters of the cognition module is performed considering the 
\textit{coupled configuration} in \autoref{fig:sep-per}. 
At this stage, the parameters of the perception module are fixed to the values obtained in the warm start, 
and only the parameters of the cognition module are identified (see steps \ref{alg:sep-per-B:start-cog-id}--\ref{alg:sep-per-B:end-cog-id} of Algorithm~\ref{alg:sep-per-B}). 
Afterward, both perception and cognition modules are warm-started by the values of their pre-identified parameters and a final identification optimization is performed on both modules (see steps \ref{alg:sep-per-B:start-final-id}--\ref{alg:sep-per-B:end-final-id} of Algorithm~\ref{alg:sep-per-B}).%
\begin{algorithm}
\caption{Model identification with warm-start - Approach B}\label{alg:sep-per-B}
\begin{algorithmic}[1]
\Require
\Statex $r$: Index of current identification run 
\Statex $n^\text{run}$: Number of identification runs
\Statex $\bm{\theta}^{\text{per-ws}}:$ Vector containing warm start of parameters of perception module obtained in Algorithm~\ref{alg:sep-per-warm-start}
\Statex $\glssymbol{params-per-run-r}$
: Vector containing identified values for parameters of perception module in run $r$
\Statex $\glssymbol{params-cog-run-r}$: Vector containing identified values for parameters of cognition module in run $r$
\Statex ${\bm{\theta}^\text{cog}_1}^*, \ldots , {\bm{\theta}^\text{cog}_{n^\text{opt}}}^*$: 
Vectors containing identified values for parameters of cognition module in the $n^\text{opt}$ runs with the minimum optimization cost until step 6
\Ensure
    \For{$r \in \{1, ..., n^\text{run}$\}} \label{alg:sep-per-B:start-cog-id}
    \State $\glssymbol{params-per-run-r} \gets \bm{\theta}^{\text{per-ws}}$ (initialization)
        \State Randomly initialize $\glssymbol{params-cog-run-r}$
        \State Optimize $\glssymbol{params-cog-run-r}$  with \textit{coupled configuration}
    \EndFor \label{alg:sep-per-B:end-cog-id}
    \State ${\bm{\theta}^\text{cog}_1}^*, \ldots , 
    {\bm{\theta}^\text{cog}_{n^\text{opt}}}^*
    \gets$  $\glssymbol{params-cog-run-r}$ of $n^\text{opt}$ runs with smallest realized values for optimization cost
    \For{$r \in \{1, \ldots, n^\text{opt}$\}} \label{alg:sep-per-B:start-final-id}
        \State $\glssymbol{params-per-run-r} \gets \bm{\theta}^{\text{per-ws}}$ (initialization)
        \State $\glssymbol{params-cog-run-r} \gets$ ${\bm{\theta}^\text{cog}_r}^*$ (initialization)
        \State Optimize $\glssymbol{params-per-run-r}$ and $\glssymbol{params-cog-run-r}$ with \textit{coupled configuration}
    \EndFor 
    \State $\glssymbol{params-per-best}, {\glssymbol{params-cog-best}} \gets \glssymbol{params-per-run-r}, \glssymbol{params-cog-run-r}$ for run $r$ with optimal optimization cost \label{alg:sep-per-B:end-final-id}
\end{algorithmic}
\end{algorithm}

On the one hand, approach~A has the advantage of being comprised of less steps, which entails less computational load. Moreover, this approach allows for a more flexible search through the parameter spaces, since all parameters are optimized together, without imposing any constraints. 

On the other hand, approach~B is first constrained with fixed parameters considered for the cognition module. 
This systematic way of exploration for the optimizer can lead to more parameters being uniquely identified, at the expense of diminishing the flexibility of the search by the optimizer.%

\subsection{Closed-loop model-based control of \glspl{sr}}\label{subsec:methods-mbc}
\begin{figure}
    \centering
    \includegraphics[width=0.48\textwidth]{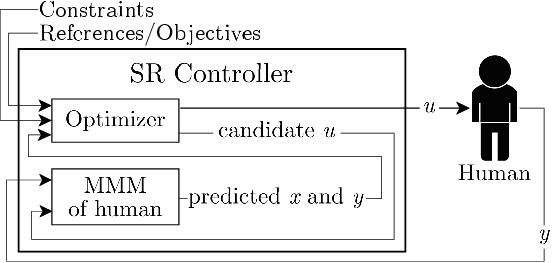}
    \caption{Block diagram of the model-based controller of the social robot using the ToM model.}
    \label{fig:mbc-control-diagram} 
\end{figure}

Once the model is mathematically and dynamically formulated, 
after identifying its parameters for a particular human, the model is embedded within a closed-loop 
control system for \glspl{sr} to be used in interactions with the human. 
\autoref{fig:mbc-control-diagram} shows the architecture of the \gls{mbc} system 
and illustrates how the dynamic \gls{tom} model, MMM, is integrated within the control loop.%

The \gls{mbc} system generates a control input $\bm{u}_{k}$, which includes all the controlled interactive actions of the robot, 
such that given criteria for the \gls{hsri} are met. 
In general, satisfying such criteria is associated with optimizing given objectives 
(e.g., maximizing the interaction time, the user engagement, or satisfaction). 
To do so, an optimizer may be used in the control loop (see \autoref{fig:mbc-control-diagram}) to 
suggest candidate control inputs. 
Note that such a controller, in general, works upon a prediction window, across which 
candidate control inputs are generated (for the current and all future time steps within the prediction window). 
In case the size of the prediction window is unity, the \gls{mbc} system looks only one step ahead 
(i.e., considers only  the impact of the current control input on the \gls{hsri}).%

In general, every \gls{hsri} is subject to various constraints that impact the actions of the \gls{sr}. 
For instance, if the \gls{sr} should offer any joint activities to a person, it should consider the 
physical and cognitive limitations of the person in making the suggestion. 
Therefore, the optimization problem of the \gls{mbc} system is in general a constrained one 
(see \autoref{fig:mbc-control-diagram}).%

Subsequently,  the candidate control inputs are fed into the dynamic \gls{tom} model, which predicts 
the mental states (shown by $x$ in the figure) of the human as a result of these stimuli (i.e., 
suggested interactive actions of the \gls{sr}), as well as the corresponding actions by the human (shown by $y$ in the figure). In case these fulfill the desired objectives and satisfy the imposed 
constraints, the candidate control inputs are selected for the \gls{sr}. 
Otherwise, the optimizer should adjust and propose new candidate control inputs, and the process is repeated.%

By following such an approach, it is possible to ensure that the \glspl{hsri} comply with 
pre-defined constraints (e.g., keeping the user engagement above a certain level) 
and that the interactive actions of the \gls{sr} result in maximizing the given criteria 
(e.g., sustaining the \glspl{hsri} as long as possible), by incorporating  the current 
(and future) mental states and actions of the user systematically within the control loop.%

\section{Case study}\label{sec:case-study}

To assess our framework, we carried out an experiment involving \glspl{hsri} with $10$ volunteer human participants. In these \glspl{hsri}, each participant solved multiple chess puzzles while interacting with a Nao robot. Each participant interacted with Nao in $3$ sessions of $45$ to $90$ minutes. 

\subsection{Experimental setup}
\label{sec:case-study-setup}

\begin{figure}
     \centering
     \begin{subfigure}[b]{0.45\textwidth}
         \centering
         \includegraphics[width=\textwidth]{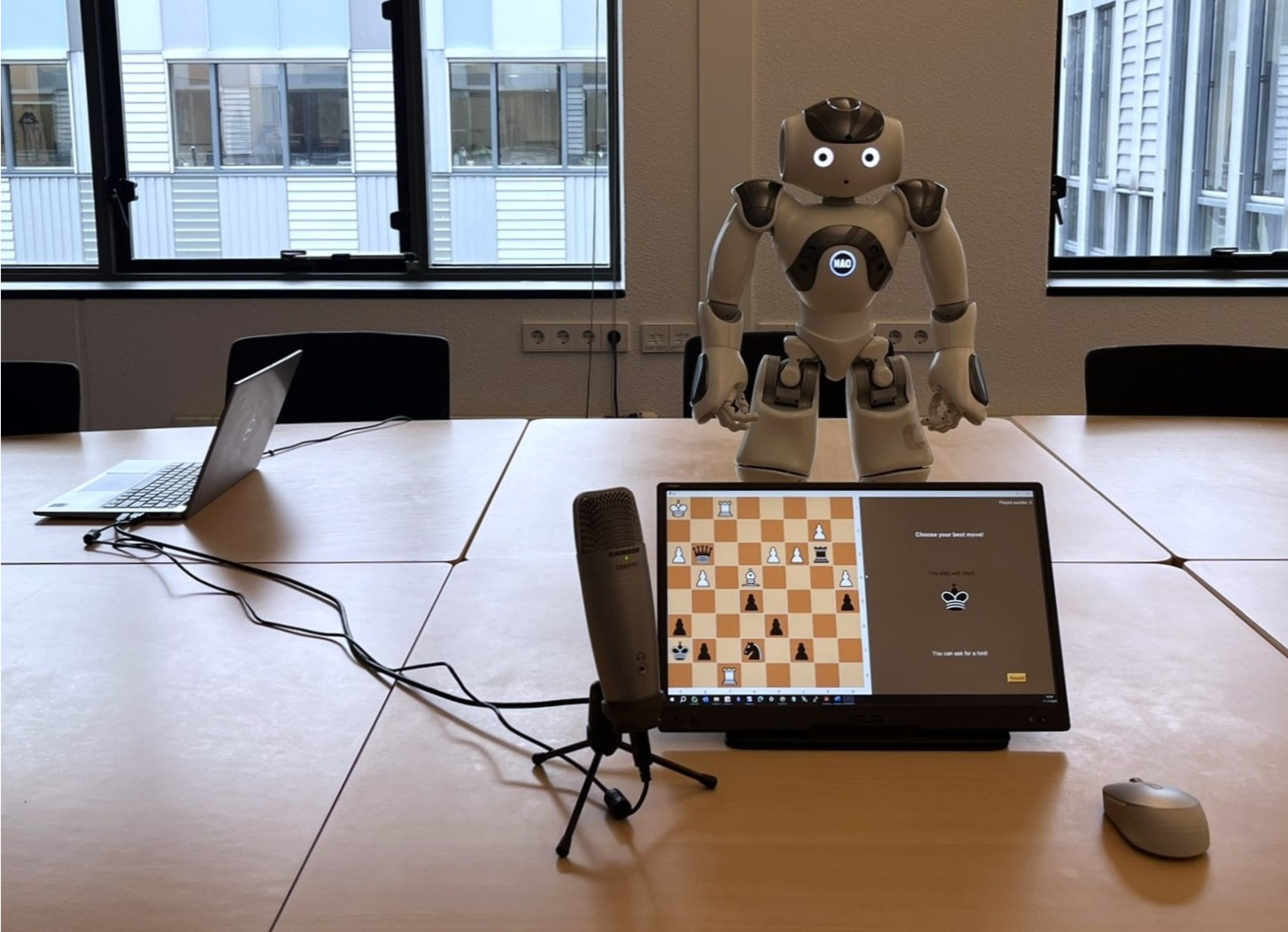}
     \end{subfigure}
     \hfill
     \begin{subfigure}[h]{0.45\textwidth}
         \centering
         \includegraphics[width=\textwidth]{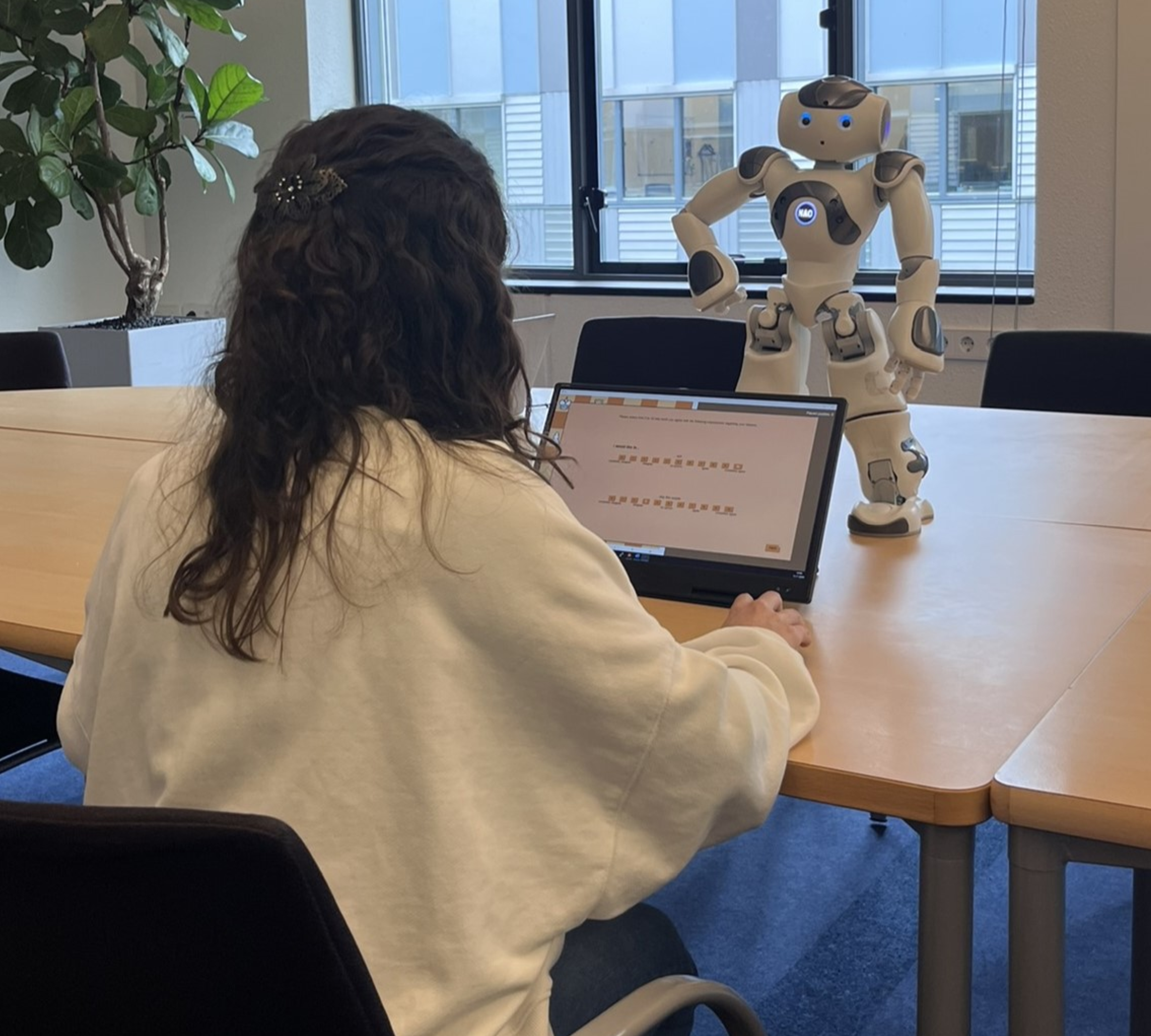}
     \end{subfigure}
         \caption{Experimental setup: The participant sat at a table in the room. The Nao robot was placed on top of that table, facing the participant. 
         A display was placed on the table between Nao and the participant. The display showed the chess puzzles 
         and the questions asked to the participant during the session. 
         A mouse and a microphone were placed on the sides of the display. The microphone and the display were both connected to a laptop that managed the \glspl{hsri} and that was connected via Wi-Fi to Nao to steer its behavior via Python.}
        \label{fig:set-up}
\end{figure}

\paragraph{Setup}
The experimental setup included a Nao robot (a programmable humanoid robot), a display, a microphone, and a mouse, as shown in \autoref{fig:set-up}. 
In each experiment session, the participant was seated at a table, with the Nao robot placed in front, facing them. A display, which was placed between the user and the robot, showed the chess puzzles, instructions, and questions to the participant during the session. A desktop microphone was placed on the side of the screen to facilitate the verbal communication of the participants with Nao, allowing more robust communication than when the microphones of the robot were used. 
Finally, a mouse was placed on the other side of the display to enable the participants to play the puzzles. 
A laptop was connected to both the Nao robot (via WiFi), the display, and the microphone (via a wired connection). This setup enabled the autonomous control of Nao through Python and facilitated data exchange during the sessions.
On the third session with each participant, a camera was placed on top of the display pointing to the participant to record their facial expressions and status during the session. This 
data was used afterward to analyze the engagement of the participants. Finally, the chess puzzles were taken from the \textit{lichess.org} database \cite{lichess_db}.

\begin{figure}
     \centering
     \begin{subfigure}[b]{0.49\textwidth}
        \centering
        \includegraphics[width=\textwidth]{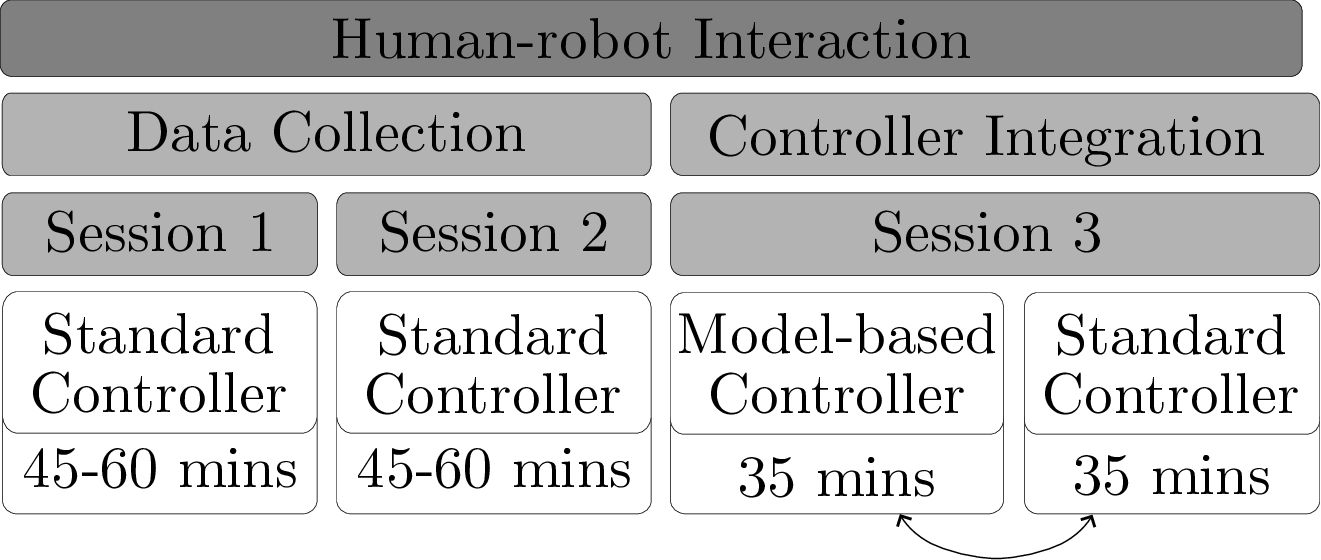}
        \caption{Description of the experiment layout, including the three sessions, their purposes, and durations.}
        \label{fig:experiment_diagram-general} 
     \end{subfigure}
     \hfill
     \begin{subfigure}[h]{0.49\textwidth}
        \centering
        \includegraphics[width=\textwidth]{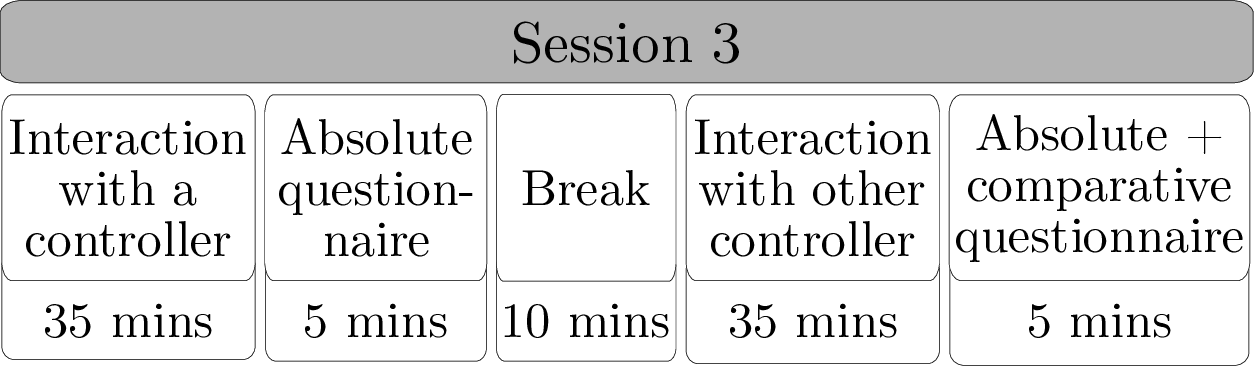}
        \caption{Detailed layout of the third session of the experiment, where the human-robot interaction using our proposed model-based steering framework, including MMM, was assessed.}
        \label{fig:experiment_diagram-session3} 
     \end{subfigure}
        \caption{Experiment layout including three separate sessions}
        \label{fig:experiment_diagrams}
\end{figure}

\paragraph{Design of interaction sessions}
The experiment with each participant consisted of 3 sessions, where the details 
are given in \autoref{fig:experiment_diagrams}.
The first two sessions were designed to collect data about the changes in the relevant mental states of the 
participant in order to identify the MMM for that participant (personalization). In the third session, the identified model was embedded in the controller of the robot, so that the robot interacted with the participant following the method described in Section~\ref{subsec:methods-mbc} and based on the personalized MMM (see \autoref{fig:experiment_diagram-general}).%

In the first two sessions, the participants were asked to solve at least $18$ chess puzzles or to continue solving puzzles for at least $45$ minutes. 
After playing $30$ puzzles or solving puzzles for $60$ minutes, the session was ended nonetheless.
The participants were in all cases allowed to terminate or quit the session any time earlier, if they wished to.  
The last session consisted of 2 interactions of $35$ minutes each. 
After each interaction, the participant was given a questionnaire to complete. Moreover, a ten-minute break was given between the two interactions (see \autoref{fig:experiment_diagram-session3}). 
In one of the two interactions in this session, the participant interacted with a version of Nao that was steered via the \gls{mbc} embedding the MMM identified for that participant. 
In the other interaction, Nao was controlled by a model-free, rule-based controller that resembled the current simplified steering systems used for \glspl{sr}. 
This controller (specified as the standard controller in Figure~\ref{fig:experiment_diagrams}) 
was also used to steer the behavior of Nao in the first two sessions designed for data collection. 
Half of the participants interacted with the standard controller in the first interaction of the third session, while the other half interacted with it in the second interaction of the third session.
For each participant, the sessions were at least one week apart from each other, and the entirety of the three sessions was conducted within one month.%

\paragraph{Execution of sessions}
Before the start of the experiment with each participant, the participant was informed about how to interact with Nao and how to play the puzzles via a briefing document. Those participants who had little or no experience with chess could request an extra document that described the basic moves and rules of chess. 
Then, the experiment started with a brief introduction given by Nao which summarized the most essential points described in the briefing document. Subsequently, the first puzzle showed up on the screen.%
The debriefing documents can be found in \referenceDataSetExperiment.

Each puzzle consisted of a sequence of $2$ to $6$ most advantageous moves that the participant had to perform correctly. To perform a move, the participant had to click on the desired piece first, and then on the square that the piece was supposed to move to. If the move was correct, the piece was placed in the new square and the next move was made by the program. 
Otherwise, the move was refused and the participant had to try again until the right move was done.  
On the right-hand side of the screen, the color that the participant had to play with and the number of
puzzles that the participant had completed so far were displayed while the participant was solving a puzzle.%

In the first and second sessions, twice per puzzle, the participant was asked to answer questions about their mental states. More specifically, the participant was asked to quantify their beliefs, goals, and emotions relevant for the situation, in a discrete range from $0$ to $10$, additionally categorized by qualitative terms from ``completely 
disagree'' to ``completely agree''. The questions showed up on the screen, and the participant 
had to reply by choosing a number using the mouse. These questions are presented in \autoref{app:questions}.%

While playing, participants could verbally ask Nao for a hint, to skip the current puzzle, or to quit the session.
Nao would then confirm the request verbally with the participant.
When participants asked Nao for the first hint about a puzzle, it provided a general tip about the objective of the puzzle (e.g., the objective was to check-mate the opponent). 
If the participant asked Nao for a second hint about the same puzzle, 
it would reveal which piece had to be moved. 
In response to requesting a third hint, if it concerned the same move, Nao would reveal the optimal move of the piece. If the hint concerned a new move, Nao would again reveal the piece to be moved.
When a participant asked to skip the current puzzle, that puzzle disappeared and the next puzzle showed up on the screen. 
At the end of each puzzle, the participant could request more difficult or easier puzzles, by replying to a question about this that appeared on the screen.
All this information was included in the briefing document that was given to the participants at the beginning of the sessions. 
This document can be found in \referenceDataSetExperiment.  

\paragraph{Participants of the experiments}
Ten volunteers aged between 25 and 35 years old participated in the experiment. The participant pool consisted of $30\%$ women and $70\%$ men.

\paragraph{Role of Nao}
Besides giving general indications about the chess puzzles to the participants and socially interacting with the participants, Nao also decided the difficulty level of the puzzles and whether to give rewards to the participants, in order to keep them engaged.%

Regarding the difficulty level of the puzzles, six levels of difficulty were defined based on the ratings of the puzzles given in the \textit{lichess.org} database \cite{lichess_db} (where the rating of a puzzle is a measure of its difficulty).
For each difficulty level, a range of the rating was considered and the puzzles with a rating in that range were included. 
The ratings that corresponded to each difficulty level considered in this case study are given in Table~\ref{tab:ratings-per-diff} of \autoref{app:difficulty}. %

To give rewards to the participants, Nao performed one of the following entertaining movements or gestures lasting around $10$ to $30$ seconds: pretending to play a guitar (posing as if the robot was holding a guitar, but there was no real guitar in the room), dancing, performing tai-chi, pretending to be an elephant, pretending to take a photo of the participant (with an imaginary camera).%

During the first two sessions, the actions of Nao were carefully scheduled as to instigate different beliefs, goals, and emotions for the participants, 
and to provide a diverse dataset to be subsequently used to identify the MMM per participant.  
%
%
Thus, to stimulate the mental states as widely as possible, and to capture the dynamic effects of increasing or decreasing the difficulty level of the puzzles when the participant was in different mental states, we distributed the difficulty levels of the puzzles to include both increasing and decreasing sequences. 
The difficulty level was kept identical for three puzzles in a row to ensure that the data gathered truly captured the effects of solving puzzles of that difficulty for the participant. Then this difficulty level would increase or decrease with a step size of $2$.  
More specifically, in the first session, the sequence of the levels of difficulties was $\{0, 2, 4, 4, 2, 0\}$, meaning that the difficulty level started at $0$ (for three consecutive puzzles), 
increased to $2$ (also for three consecutive puzzles), and then to $4$. 
After playing two groups of three puzzles of difficulty $4$, the difficulty level decreased to $2$, and then to $0$. 
In the second session, the sequence of the levels of difficulties was $\{5, 3, 1, 1, 3, 5\}$. 
In these sessions, if the number of puzzles that the participants solved surpassed $18$ puzzles, 
then the difficulty level of puzzle number $18+n^\text{puzzle}$, 
with $n^\text{puzzle}$ an integer, was the same as the difficulty level of puzzle $n^\text{puzzle}$  (taking into account that the maximum number of puzzles offered per session was $30$).  
See \autoref{tab:diff-per-interactions}  of \autoref{app:difficulty} for the detailed distribution of the difficulty levels in the first two sessions of the experiments.%

As for giving rewards to the participants, in the first two sessions, Nao randomly performed one 
celebratory movement or gesture for each set of three puzzles with the same difficulty level. 
Note that the main objective of this design choice was to capture the dynamic effects of giving a generally encouraging/entertaining reward on the mental states of the participants, rather than the specific design of such rewards.%

\subsection{Model-based controller for Nao}
\label{sec:MBC_controller}
The \gls{mbc} steering system of Nao used in the third session of this case study had to decide, at the end of each puzzle, about the level of difficulty of the next puzzle and about whether to give a reward to the participant. 
This decision had to optimize the mental states of the participant according to a cost function that depended on the model identified for each participant. Next we describe the details for designing the \gls{mbc} steering system of Nao.

\subsubsection{MMM for participants}
\label{sec:case-study-model}

One of the crucial elements of an \gls{mbc} steering system is a model of the  
process (i.e., the perception, cognition, and \dm\space of the participants) that should be impacted in desired ways by the designed steering system. 
This section describes the mathematical modeling of the perception, cognition, and \dm\space modules for the participants of this case-study, based on the MMM detailed in Section~\ref{sec:MMM_leverage}. 
A graphical representation of each module is given in \autoref{app:model}. 

\paragraph{Perception module}

\begin{table*}[]
\caption{Concepts that represent the \rld\ and \pd\ of the perception module for the case study.} 
\label{tab:list-of-variables-perception}
\begin{center}
\begin{tabularx}{\linewidth}{lXll}
\toprule
\cellcolor{gray!20}Real life data & \cellcolor{gray!20}Perceived data (corresponding to the \rld\ in the same row of this table) & \cellcolor{gray!20}Concept                          & \cellcolor{gray!20}Type  \\ 
\midrule
$\realLifeData{1}(\kp)$ & $\perceivedData{1}(\kp)$ & Puzzle difficulty level     & Integer \\
$\realLifeData{2}(\kp)$ & $\perceivedData{2}(\kp)$ & Number of hints       & Integer \\
$\realLifeData{3}(\kp)$ & $\perceivedData{3}(\kp)$ & Number of wrong attempts  & Integer \\
$\realLifeData{4}(\kp)$ & $\perceivedData{4}(\kp)$ & Time to solve a puzzle    & Continuous \\
$\realLifeData{5}(\kp)$ & $\perceivedData{5}(\kp)$ & Skipped puzzle?        & Boolean \\
$\realLifeData{6}(\kp)$ & $\perceivedData{6}(k)$ & Reward given?          & Boolean \\
\bottomrule
\end{tabularx}
\end{center}
\end{table*}

\begin{table*}[]
\caption{Concepts that represent the \rpk, \pk, beliefs, goals, emotions, and biases related to the cognitive module for the case study. Note that the number of the relevant beliefs, goals, emotions, and biases involved in the experiment are, respectively, $2$, $4$, $2$, and $1$.} 
\label{tab:list-of-variables-cognition}
\begin{center}
\begin{tabularx}{\linewidth}{lXXl}
\toprule
\cellcolor{gray!20}{Rationally} & \cellcolor{gray!20}{Perceived knowledge (corres-} & \cellcolor{gray!20}{Beliefs (corresponding to} & \cellcolor{gray!20}Concept     
\\ 
\cellcolor{gray!20}{perceived} & \cellcolor{gray!20}{ponding to the rationally} & \cellcolor{gray!20}{the \pk\ in} &   \cellcolor{gray!20}   
\\ 
\cellcolor{gray!20}{knowledge} & \cellcolor{gray!20}{perceived knowledge in the same row of this table)} & \cellcolor{gray!20}{the same row of this table)} &   \cellcolor{gray!20}   
\\ 
\midrule
{$\rationallyPerceivedKnowledge{1} (k)$} & {$\perceivedKnowledge{1}(k)$} & 
{$\belief{1} (k)$} &  The puzzle is difficult \\ 
{$\rationallyPerceivedKnowledge{2} (k)$} & {$\perceivedKnowledge{2}(k)$} & 
{$\belief{2} (k)$} &  Nao offered a reward \\
\toprule
\multicolumn{3}{l}{\cellcolor{gray!20}Goals} & \cellcolor{gray!20}Concept     
\\ 
\midrule
\multicolumn{3}{l}{$\goal{1} (k)$} & Quit the session\\
\multicolumn{3}{l}{$\goal{2} (k)$} & Skip the puzzle\\
\multicolumn{3}{l}{$\goal{3} (k)$} & Get help\\
\multicolumn{3}{l}{$\goal{4} (k)$} & Change difficulty\\
\toprule
\multicolumn{3}{l}{\cellcolor{gray!20}Emotions} & \cellcolor{gray!20}Concept     
\\ 
\midrule
\multicolumn{3}{l}{$\emotion{1} (k)$} & Boredom\\
\multicolumn{3}{l}{$\emotion{2} (k)$} & Frustration\\
\toprule
\multicolumn{3}{l}{\cellcolor{gray!20}Bias} & \cellcolor{gray!20}Concept     
\\ 
\midrule
\multicolumn{3}{l}{$\bias (k)$} & The puzzle is difficult\\
\bottomrule
\end{tabularx}
\end{center}
\end{table*}


The perception module, as explained before (see \autoref{fig:perception-module}), is in general composed of perceptual access 
and rational reasoning sub-processes. 
\autoref{tab:list-of-variables-perception} shows the pieces of \rld\space and \pd\space that are relevant for 
and used in the case study. The list of the pieces of \rpk\space is given in \autoref{tab:list-of-variables-cognition}. 
In this case study, due to the nature and setup of the experiments, it is sensible to assume that the 
participants perfectly observe all relevant pieces of \rld.
Thus, based on \eqref{eq:perception-general-pa}, 
for all perception time steps $\kp$, for the perceptual access we have $\perceivedData{i} (\kp) = \realLifeData{i}(\kp)$, where $\perceivedData{i} $ is a piece of perceived data (i.e., an output from the perceptual access sub-process) 
and $\realLifeData{i}(\kp)$ is its corresponding input \rld. 
As given in \autoref{tab:list-of-variables-perception}, for this case study, $i = 1, \ldots, 6$.%
%

To model the rational reasoning sub-process, which is formulated via \eqref{eq:perception-general-rr} in MMM, 
the relationships between all pieces of the \pd\space $\perceivedData{i} (\kp)$ for $i=1, \ldots,6$ 
(i.e., the inputs to the sub-process) and the corresponding piece of \rpk\space $y_j^{\text{RR}}(\kp)$
(i.e., the outputs of the sub-process) 
should be defined by providing relevant expressions for function $f^{\text{RR}}_{ij}(\cdot)$ in \eqref{eq:perception-general-rr}. 
\autoref{fig:perception_module} illustrates the pieces of \pd\space that influence each \rpk, where 
$\rationallyPerceivedKnowledge{1}(\kp)$ in \autoref{tab:list-of-variables-cognition} is influenced by $\perceivedData{1} (\kp), \ldots,\perceivedData{5} (\kp)$ given in \autoref{tab:list-of-variables-perception}, 
and $\rationallyPerceivedKnowledge{1}(\kp)$ is influenced by $\perceivedData{6} (\kp)$.%

In order to model function $f^{\text{RR}}_{ij}(\cdot)$ 
to mathematically represent the processes that are illustrated in  \autoref{fig:perception_module}, 
three types of mathematical expressions were considered, i.e., affine \eqref{eq:perception-prop}, exponential \eqref{eq:perception-exp}, and boolean \eqref{eq:perception-bool}. The choice of each expression depended on the nature of the input variable
$\perceivedData{i} \left(\kp\right)$ as is explained further when the particular use cases of these functions are discussed.  The expressions for these functions are:
\begin{subequations}
\label{eq:perception}
\begin{align}
\begin{split}
\label{eq:perception-prop}
f^\text{RR,F}_{ij} & \left(\perceivedData{i} \left(\kp\right); \theta_{ij,0},\theta_{ij,1}\right) = \\
& = \theta_{ij,0} \perceivedData{i} \left(\kp\right) + \theta_{ij,1}
\end{split} \\
\begin{split}
\label{eq:perception-exp}
f^\text{RR,E}_{ij}  & \left(\perceivedData{i} \left(\kp\right); \theta_{ij,0},\theta_{ij,1}\right) = \\
& = \theta_{ij,1}  \exp{\left(\theta_{ij,0}  \perceivedData{i} \left(\kp\right)\right)} + 1
\end{split} \\
 \begin{split}
 \label{eq:perception-bool}
 f^\text{RR,B}_{ij}  & \left(\perceivedData{i} \left(\kp\right); \theta_{ij,0},\theta_{ij,1}\right) = \\
 & = \begin{dcases}
     \theta_{ij,0} \text{,} & \perceivedData{i} \left(\kp\right)=0 \\
     \theta_{ij,1} \text{,} & \perceivedData{i} \left(\kp\right)=1
 \end{dcases} 
\end{split}
\end{align}
\end{subequations}

We have selected the affine relationship \eqref{eq:perception-prop} to estimate the impact of the inputs 
that are integer, with an upper bound. These inputs include \textit{puzzle difficulty level}, $\perceivedData{1} (\kp)$, and \textit{number of hints}, $\perceivedData{2} (\kp)$, both being integers with an upper bound of, respectively, $5$ and $1+2 n^\text{moves}$ where $n^\text{moves}$ is an integer that represents the number of moves in the puzzle.%
%
For the inputs of the rational reasoning sub-process that are integers but have no upper bound, 
i.e., for \textit{number of wrong attempts}, $\perceivedData{3} (\kp)$, and \textit{time to solve a puzzle}, $\perceivedData{4} (\kp)$, the impact on the output of the sub-process was modeled using an exponential function, as given in \eqref{eq:perception-exp}.
%
Finally, for those inputs to the sub-process that are of a Boolean nature, i.e., for \textit{skipped puzzle?}, $\perceivedData{4} (\kp)$, and \textit{reward given?}, $\perceivedData{5} (\kp)$, the impact was modeled using  \eqref{eq:perception-bool}.%

The first output $\rationallyPerceivedKnowledge{1}(\kp)$ of the rational reasoning sub-process, i.e., \textit{the puzzle is difficult}, is generated by all first $5$ pieces of inputs from the \pd\ indicated in \autoref{tab:list-of-variables-perception}, i.e., by $\perceivedData{1} (\kp), \ldots, \perceivedData{5} (\kp)$, 
as is shown in \autoref{fig:perception_module}.  
The second output $\rationallyPerceivedKnowledge{2}(\kp)$, i.e., 
\textit{Nao offered a reward}, is generated solely by $\perceivedData{6} (\kp)$, i.e., by the perceive data \textit{reward given?} (see \autoref{tab:list-of-variables-perception}). 
Using \eqref{eq:perception-general-rr} and \eqref{eq:perception}, we have: 
\begin{subequations}
\begin{align}
\begin{split}
\label{eq:perception-difficulty}
\rationallyPerceivedKnowledge{1} (\kp)
& = 
\theta_{{11},0}  y^\text{PA}_1 (\kp) + \theta_{{11},1} + \\
 &  \theta_{{21},0} y^\text{PA}_2 (\kp) + \theta_{{21},1} +\\
 & \theta_{{31},1} \exp\left({\theta_{{31},0} \perceivedData{3}} (\kp)\right) + 1) +\\
 & \theta_{{41},0} \exp\left({\theta_{{41},1} \perceivedData{4}} (\kp)\right) + 1) +\\ 
& f^\text{RR,B}_{51}  \left(\perceivedData{5} \left(\kp\right); \theta_{51,0},\theta_{51,1}\right)
\end{split} \\
\begin{split}
\label{eq:perception-var-reward}
\rationallyPerceivedKnowledge{2} (\kp)
& = f^\text{RR,B}_{61}  \left(\perceivedData{6} \left(\kp\right); \theta_{61,0},\theta_{61,1}\right)
\end{split}
\end{align}
\end{subequations}

\begin{remark}
    Note that in \eqref{eq:perception-difficulty} the term $\theta_{{11},1} + \theta_{{21},1} +  f^\text{RR,B}_{51}  \left(\perceivedData{5} \left(\kp\right); \theta_{51,0},\theta_{51,1}\right)$ may be identified as one single parameter. 
\end{remark}

\paragraph{Cognition module}

The cognition module (see \autoref{fig:cognition-module}) is composed of three main state variables, 
i.e., beliefs, goals, and emotions. 
In the setup of this case study, the number of beliefs, goals, and emotions involved is, respectively, 
$2$, $4$, and $2$, where the corresponding concepts have been given in \autoref{tab:list-of-variables-cognition}. The realization of these variables are real values which can vary between $-1$ and 1. 
\autoref{fig:cognitive_module_complete} and \autoref{fig:cognitive_module_simple} show the structure of the cognitive module, including variables and linkages.%
\autoref{fig:cognitive_module_complete} corresponds to the complete model used in the first and second sessions. Following the strategy described in Section \ref{sec:methods-identification}, a simplified version of the previous model was used to integrate the model-based control session. \autoref{fig:cognitive_module_simple} represents this simplified model. 

To generate the perceived knowledge, \eqref{eq:update-FCM} is used. The first term of this equation is null (i.e., $f_i(x_i(k))=0$), and $z_{\ell i}(l)=1$. As shown in \autoref{fig:cognition-module}, the second term of this equation reflects the influence of the biases. 

In order to update the beliefs, goals, and emotions, \eqref{eq:update-FCM}--\eqref{eq:Definition_weight} are used, with weight $w_{ii}$ set to $0$ for beliefs and to $0.9$ for both goals and emotions. The weights $w_{ji}(k)$ are given by:
\begin{equation}
\label{eq:sw-case-study}
w_{ji} (k) = 
\begin{cases}
    w^{^-}_{ji}, & \glssymbol{state-variable}_j(k) \leq 0 \\
    w^{^+}_{ji}, & \glssymbol{state-variable}_j(k) > 0
\end{cases} 
\end{equation}
where the parameter values $w^{^+}_{ji}$ and $w^{^-}_{ji}$ should be identified for each participant. 

\begin{remark}
\label{rem:hypothesis_EGB_frequencies}
To assess the validity of the hypothesis that the goals and emotions are updated with a lower frequency than the beliefs, as stated in Section~\ref{sec:MMM_leverage}, in addition to considering the values indicated above for $w_{ii}$, an additional identification procedure for the MMM has been conducted for all participants of the case study using $w_{ii}=0$ for both the beliefs and the goals. The estimation errors have been shown and compared in Section~\ref{sec:results}, and the validity of the hypothesis has been discussed there. 
\end{remark}
\begin{remark}
Since the biases are auxiliary variables used to describe the effect of emotions on new beliefs, we represent them without dynamics.   
\end{remark}

\paragraph{Decision-making module}

The \dm\ module (see \autoref{fig:decision-making}) is composed of two main sub-processes, 
rational intention selection and rational action selection. 
\autoref{tab:list-of-variables-dm} includes the concepts that define the intentions and actions relevant for the decision-making module in this case study. 

Regarding the rational intention selection, the intensity of each intention $\intention{i}$ is given by \eqref{eq:dm-activation-functions}.
In the context of the case study, each intention is only influenced by one goal and is not influenced by any belief, yielding a particular case of \eqref{eq:dm-activation-functions}. 
\autoref{tab:list-of-variables-dm} lists the goal that influences each intention, where the $4$ goals, described earlier in this section, result in $5$ possible intentions. This information is also graphically presented in \autoref{fig:dm_module}.%

As for the action selection, we assume that each action is performed if the intention corresponding to the same concept is positive (and not performed otherwise). Hence, the rational action selection is mathematically represented as:
\begin{equation}
    \action{i} (k) = 
    \begin{cases}
        0 \text{,} & \intention{i} (k) \leq 0 \\
        1 \text{,} & \intention{i} (k) > 0
    \end{cases} 
\end{equation}

\begin{table*}[]
\caption{Concepts that represent the intentions and actions of the \dm\ for the case study.} 
\label{tab:list-of-variables-dm}
\begin{center}
\begin{tabularx}{\linewidth}{lXll}
\toprule
\cellcolor{gray!20}Intention & \cellcolor{gray!20}Action & \cellcolor{gray!20}Concept & \cellcolor{gray!20}Cognitive variable that influences the intention  \\ 
\midrule
$\intention{1} $ & $\action{1} $ & Quit the game                     & $\goal{1} (k)$: Quit the game \\ 
$\intention{2} $ & $\action{2} $ & Skip the puzzle                   & $\goal{2} (k)$: Skip the puzzle \\
$\intention{3} $ & $\action{3} $ & Ask for help                      & $\goal{3} (k)$: Get help; \\ 
$\intention{4} $ & $\action{4} $ & Ask for an easier puzzle          & $\goal{4} (k)$: Change difficulty \\ 
$\intention{5} $ & $\action{5} $ & Ask for a more difficult puzzle   & $\goal{4} (k)$: Change difficulty \\ 
\bottomrule
\end{tabularx}
\end{center}
\end{table*}

\paragraph{Measurement frequency}

The \gls{mbc} should determine whether or not the participant receives a reward from Nao, 
as well as the difficulty level of the next puzzle, every time a puzzle is finished. 
Since the measurements of the cognitive variables are gathered by asking the participants about their mental states 
(see Appendix~\ref{app:questions}), there needs to be a trade-off between collecting a sufficient number of measurements and not disturbing or distracting the participants, as it may affect the reliability of the measurements. 
Therefore, to balance the number of questions asked, 
they were posed twice per puzzle and after the participant requested a hint.

Every time a measurement of the mental states was received through the feedback of the participant, 
the inputs given in \autoref{tab:list-of-variables-perception} were also collected. These inputs were directly accessible in the code. Then, the estimates for the perception module of MMM were updated. 
The cognition module, as explained in Section~\ref{sec:MMM_leverage}, may be updated according to a frequency different from the frequency of capturing the measurements. 
In this case study, the variables of the cognition module were updated two times in between every two consecutive measurements.%
%

\subsubsection{Model identification}
\label{sec:case-study-model-id}

Preliminary \gls{hsri} were carried out with two volunteer participants, 
where in addition to the concepts presented in Section~\ref{sec:case-study-model} (see \autoref{tab:list-of-variables-cognition}), the preliminary model included two extra concepts for the belief, i.e., \textit{Nao hinders solving the puzzle} and \textit{participant made progress}, one for the goal, i.e., \textit{get a reward}, and one for the emotion, i.e.,  \textit{happiness}. Moreover, in addition to determining the difficulty level of the next puzzle and whether or not 
to offer a reward to the participant,  Nao had a choice to hinder the participant by purposely giving incorrect clues or to help the participant.  Hindering was hypothesized to induce frustration in the participants, generating more diverse data and improving the identification process.
Note that these two volunteer participants differed from the $10$ participants of the experiments, and their data has not been included in the results. Their participation aimed to assess the designed sessions by repeating the experiments multiple times, collecting feedback, and assessing the estimations. By doing so, the design of the sessions was enhanced as much as possible before involving the participants. 

It was found out that, with the data collected in two sessions lasting $45$ to $60$ minutes per participant, it was not possible to identify all parameters unequivocally, deploying the identification and assessment methods presented in Section~\ref{sec:methods-identification}. Performing a third session with each participant (i.e., extending the training dataset) slightly improved the number of parameters that were unequivocally identifiable, but did not completely address the issue. 
It was concluded that the control input concerning Nao yielding false hints and its corresponding belief could create confusion, especially for participants with no or limited experience with chess playing. 
This introduces undesirable noise into the data, hindering the identification process. 
By analyzing the data collected in the preliminary sessions, the belief \textit{participant made progress}, the goal \textit{get a reward}, and the emotion \textit{happiness} showed to be redundant. Furthermore, the other concepts for these variables were more relevant to be tracked and optimized by Nao. 
Consequently, the model was simplified according to the approach that has been presented in Section~\ref{sec:methods-identification}.

After the preliminary phase, the model described in Section~\ref{sec:case-study-model}, which we refer to as \textit{complete model} and is shown in \autoref{fig:cognitive_module_complete}, was obtained. 
This \textit{complete model} was identified for both preliminary participants in order to assess its accuracy in tracking their mental states.
Then, two extra simplifications were designed to reduce the model that was embedded in the \gls{mbc} used in the last session. 
The first simplification was the removal of the goals \textit{get help} and  \textit{change difficulty} (see
\autoref{fig:cognitive_module_simple} for a representation of the \textit{simplified model}), 
as these goals did not directly reflect the quality of the interactions (and, consequently, did not need to be optimized), nor did they influence the other mental states.  
For the ten volunteer users, the choice to embed the complete or the simplified model in the \gls{mbc} was done per user, by selecting the model that achieved a smaller validation error and 
that corresponded to a larger number of unequivocally identifiable parameters. 
The second simplification was the replacement of the scheduled weights in \eqref{eq:sw-case-study} by just one scalar parameter (i.e., $w_{ji}(k)=w^{^-}_{ji}=w^{^+}_{ji}$). This second simplification was considered only for participants whose simplified model could not be fully unequivocally identified.

Note that the weight parameters $w_{ii}$ in \eqref{eq:sw-case-study} were set to $0.9$ 
whenever $x_i(k)$ represented a goal or an emotion. 
For the participants for whom setting $w_{ii}=0.9$ did not result in a satisfactory optimization cost, $w_{ii}$ was gradually reduced with a step of $0.1$ until the optimization cost was acceptable. 
Finally, to facilitate and improve the model identification, the inputs of the \rld\space were normalized per participant between 0 and 1. 

\subsubsection{Optimizer}
The integration of the MMM into an \gls{mbc} is based on \autoref{fig:mbc-control-diagram}. 
The control input is composed of two variables, where the first one, i.e., the difficulty level of 
the next puzzle, is an integer variable with $6$ possible realizations (see \autoref{app:difficulty}) and 
the second one, i.e., \textit{reward given?}, is a Boolean with $2$ possible realizations ($1$ when reward is given, and $0$ otherwise). 
Therefore, $12$ possible combinations, and $12$ possible control inputs, exist. 

A cost function was formulated based on the belief, goals, and emotions 
that are predicted at time step $k$ for the upcoming time step $k+1$ by the MMM, 
when the current state vector $\bm{x}(k) = [\belief{1} (k), \belief{2} (k), \goal{1} (k),\goal{2} (k), 
\goal{3} (k), \goal{4} (k), \emotion{1} (k),\emotion{2} (k)]^{\top}$ is measured:
\begin{equation}
\label{eq:cost-function}
J({\bm{x}}(k)) = |\belief{1} (k + 1)| + w^{\text{g}}\sum_{i = 1}^2 \goal{i} (k + 1) + w^{\text{e}} \sum_{i = 1}^2 \emotion{i} (k+1) 
\end{equation}
The estimation of the upcoming state variables based on the current measured states has been detailed in Section~\ref{sec:case-study-model}.%

The formulation of \eqref{eq:cost-function} balances a trade-off in minimizing the cognitive variables whose rise will negatively 
impact the participants, including the absolute value of the first element of the belief, i.e., \textit{the puzzle is difficult} (see \autoref{tab:list-of-variables-cognition}), 
the first two elements of the goal (i.e., \textit{quit the session} and \textit{skip the puzzle}), 
and both elements of the emotion (i.e., \textit{boredom} and \textit{frustration}). The lower the value of the mentioned goals and emotions is, the better for the \gls{hsri}. As for the belief, since the minimum and maximum values of this variable correspond to the belief that the puzzle is too easy or too difficult, respectively, it is ideal to keep this belief as neutral as possible (i.e., as close to $0$ as possible).
The parameters $ w^{\text{g}}$ and $ w^{\text{e}}$ weigh these concepts relatively.  
In this case study, we gave an equal level of importance to all these concepts 
and tuned them as $w^{\text{g}} = w^{\text{e}} = 1$.
The control input that yields in the lowest value of the cost function given by \eqref{eq:cost-function} is selected by the \gls{mbc} for time step $k$.%

\section{Results and discussions}
\label{sec:results}

Next, we present the results and discussion on identifying the MMM for the participants and using the identified models in \glspl{mbc} to steer behavior of Nao in the final interaction session.

\subsection{Results for model identification}
In order to assess the accuracy of the model in estimating the 
mental states, we divided the data points of the first two sessions into training data ($67\%$) and test data ($33\%$). Thus, the training data was used to identify the model and the test data to assess it.
For this purpose, we used the complete model composed of the variables shown in Tables~\ref{tab:list-of-variables-perception},~\ref{tab:list-of-variables-cognition}, and \ref{tab:list-of-variables-dm}, and displayed in \autoref{fig:cognitive_module_complete}. 
Note that in general, the preferences and personality traits, which remain constant in the short term 
(e.g., during the three interaction sessions), regulate the variations in the mental states (see \cite{paper1} for more details). 
In this paper, due to the nature of the \glspl{hsri}, the overall interest of the participant in playing chess (which 
is a general preference) may regulate the speed for reaching the goal ``quit the game''. Moreover, how focused and confident a participant is in general (as a trait), as illustrated in \autoref{fig:cognitive_module_complete}, impacts the evolution of 
the emotions ``boredom'' and ``frustration'', respectively. Thus, these fixed parameters have been included in  
MMM and are identified for the participants.%

The average MSE in the estimation of the mental states is presented in  \autoref{tab:MSE} for when the goals and emotions are updated with a frequency lower than the frequency of updating the beliefs, and when the goals and emotions are updated with the same frequency as the beliefs (see Remark~\ref{rem:hypothesis_EGB_frequencies}, Section~\ref{sec:case-study-model}). These average MSE values are presented for three cases concerning the identification procedure: 
(1) \textbf{Conventional}: When the coupled configuration (see \autoref{fig:sep-per}) is used to identify 
the perception and cognition modules simultaneously, using a multi-start optimization but no warm start. 
(2) \textbf{Approach A}: When the perception and cognition modules are identified following Algorithms~\ref{alg:sep-per-warm-start} and \ref{alg:sep-per-A} of Section~\ref{sec:methods-identification}. 
(3) \textbf{Approach B}: When the perception and cognition modules are identified following Algorithms~\ref{alg:sep-per-warm-start} and \ref{alg:sep-per-B} of Section~\ref{sec:methods-identification}.%

The average MSE values achieved for both the training and test data sets are very satisfactory, all 
presenting a value under $0.1$, where the maximum possible value for the MSE is $4$. 
These values indicate the excellent capacity of MMM to estimate the invisible cognitive procedures of various  
participants in a personalized way for the given \glspl{hsri}. Furthermore, they demonstrate that the identification was carried out successfully. 
Furthermore, the hypothesis that the goals and emotions should be updated with a lower frequency than the beliefs is supported by the results, where applying this hypothesis results in a decrease in the average MSE 
of around $16\%$ and $11\%$ 
for the training and test data sets, respectively. 

Finally, the identification of the perception and cognition following approaches A and B improves the percentage of parameters that are unequivocally identified when compared with the conventional approach, as shown in \autoref{tab:percentage-ided-parameters}. Therefore, these two approaches contribute to making the model more interpretable, while not impacting the accuracy of the estimations made by the identified models, as shown in  \autoref{tab:MSE}.

Next, we present the results obtained when using the \gls{mbc} for behavioral control of Nao.%

\begin{table}[]
\caption{Average MSE obtained when estimating the mental states (emotions, goals, beliefs) using the training and the test data sets in the model identification procedure. The error is normalized over all the mental states (beliefs, goals, and emotions) and time steps. Given that the mental states are bounded in $[-1, 1]$, the maximum MSE is 4. The values presented are an average over all the participants.}
\label{tab:MSE}
\centering
\begin{tabular}{llccc}
\toprule
\multirow{3}{*}{\begin{tabular}[c]{@{}l@{}}Data\\ Set\end{tabular}} & \multirow{3}{*}{\begin{tabular}[c]{@{}l@{}}Different\\ frequen-\\cies\footnotemark[1]\end{tabular}} & \multicolumn{3}{c}{\begin{tabular}[c]{@{}c@{}}Identification of \\perception and cognition\end{tabular}}                                                   \\ \cmidrule{3-5}
                                            &                                       
                                                & \multirow{2}{*}{\begin{tabular}[c]{@{}c@{}}Conven-\\tional\end{tabular}}
                                                    & \multirow{2}{*}{\begin{tabular}[c]{@{}c@{}}Approach\footnotemark[2]\\ A\end{tabular}} & \multirow{2}{*}{\begin{tabular}[c]{@{}c@{}}Approach\footnotemark[2]\\ B\end{tabular}} \\
                                                    &                                                                                      &                           &                                                                       &                                                                       \\
\midrule
\multirow{2}{*}{Training}                                              & No                                                                                   & 0.063                     & 0.064                                                                 & 0.065                                                                 \\
                                                                    & Yes                                                                                  & 0.053                     & 0.053                                                                 & 0.053                                                                 \\
\multirow{2}{*}{Test}                                               & No                                                                                   & 0.075                     & 0.075                                                                 & 0.082                                                                 \\
                                                                    & Yes                                                                                  & 0.067                     & 0.067                                                                 & 0.067                                                                
              
                                                                    \\ \bottomrule
\end{tabular}
\footnotetext[1]{Goals and emotions are once updated with a lower frequency than beliefs and once with the same frequency, as explained in Remark~\ref{rem:hypothesis_EGB_frequencies}, Section \ref{sec:case-study-model}}
\footnotetext[2]{Approaches A and B are explained in Section \ref{sec:methods-identification}}
\end{table}

\begin{table}[]
\caption{Percentage of parameters unequivocally identified during the optimization process with the different identification approaches. The values presented are an average over all the participants.}
\label{tab:percentage-ided-parameters}
\centering
\begin{tabular}{@{}lccc@{}}
\toprule
\multirow{3}{*}{\begin{tabular}[c]{@{}l@{}}Different\\ frequencies\end{tabular}} & \multicolumn{3}{c}{Identification of  perception and cognition}                           \\ \cmidrule(l){2-4} 
                                                                                 & \multirow{2}{*}{Conventional} & \multirow{2}{*}{Approach A} & \multirow{2}{*}{Approach B} \\
                                                                                 &                               &                             &                             \\ \midrule
No                                                                               & $59.44\%$                         & $76.94\%$                       & $80.56\%$                      \\
Yes                                                                              & $73.61\%$                         & $96.39\%$                       & $97.78\%$  \\ 
\bottomrule
\end{tabular}
\end{table}

\subsection{Results for model-based controller}

\begin{figure*}[]
    \centering
    \includegraphics[width=\textwidth]{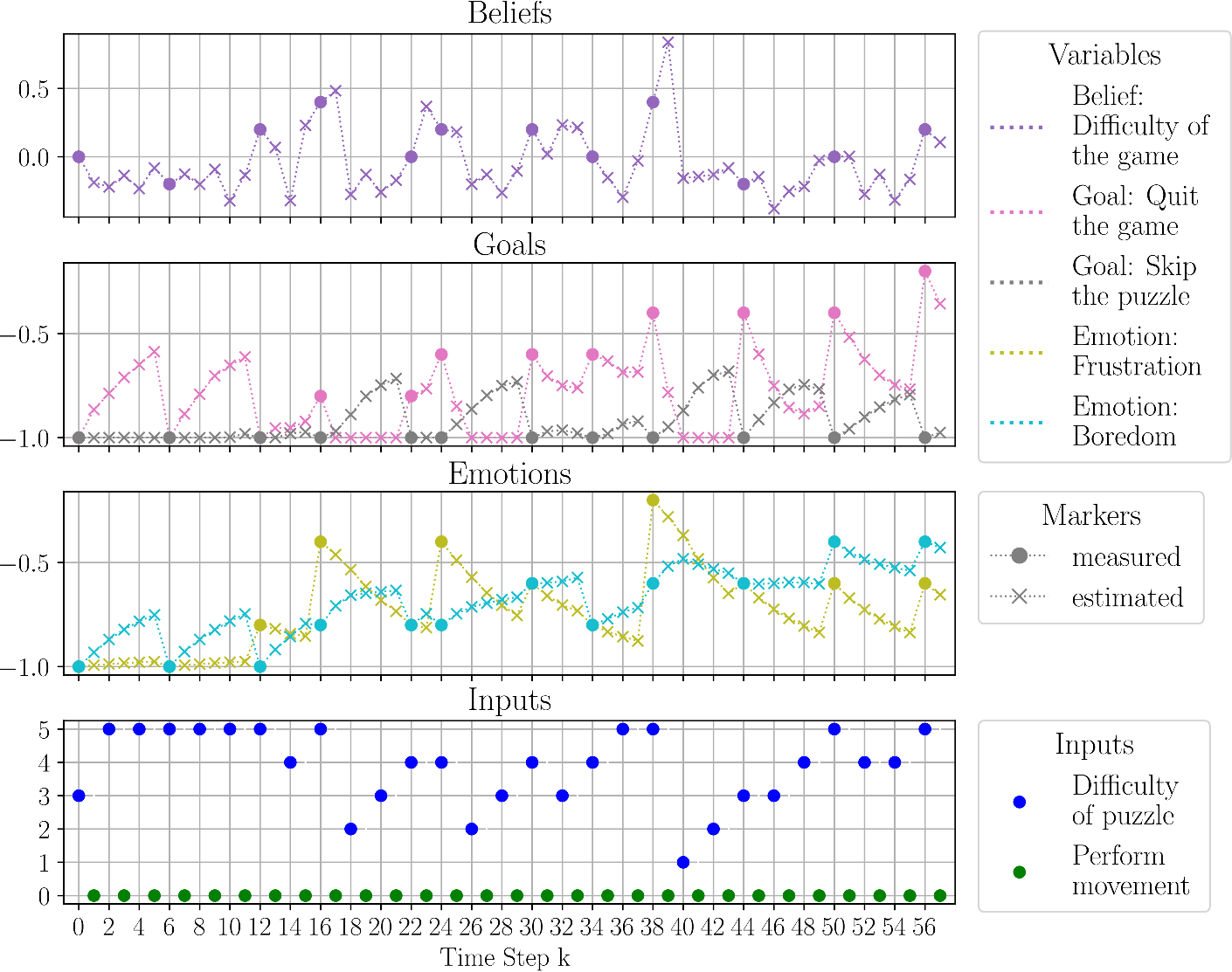}
    \caption{Values corresponding to the mental states and inputs (i.e., decision variables of the controller) for one participant over time. 
    The vertical lines correspond to the start of a new puzzle.
    After the introduction of a puzzle, two time steps were considered, such that the first time step was in the middle of playing the puzzle and the second time step coincided with the start of the next puzzle. 
    Around time step $k=16$, the participant suddenly became frustrated, as is deduced from the plot for emotions. Thus, in the next time step, the controller lowered the puzzle difficulty level, which, over time (i.e., by reaching time step $k=30$), led to a decrease in the frustration. The same happens again at time step $k=38$. Moreover, when the participant presented the same level of frustration with a higher level of boredom (see time steps $k=44$ and $k=40$), the difficulty level of the puzzle for the next time step was chosen to be higher by the controller. 
    The controller effectively prevents the engagement of the participant from increasing excessively during interactions by appropriately responding to their mental states
    As for the reward, the controller has inferred that the participant preferred not to receive rewarding movements from Nao.}
    \label{fig:individual-case}
\end{figure*}

\begin{figure*}[]
    \centering
    \includegraphics[width=\textwidth]{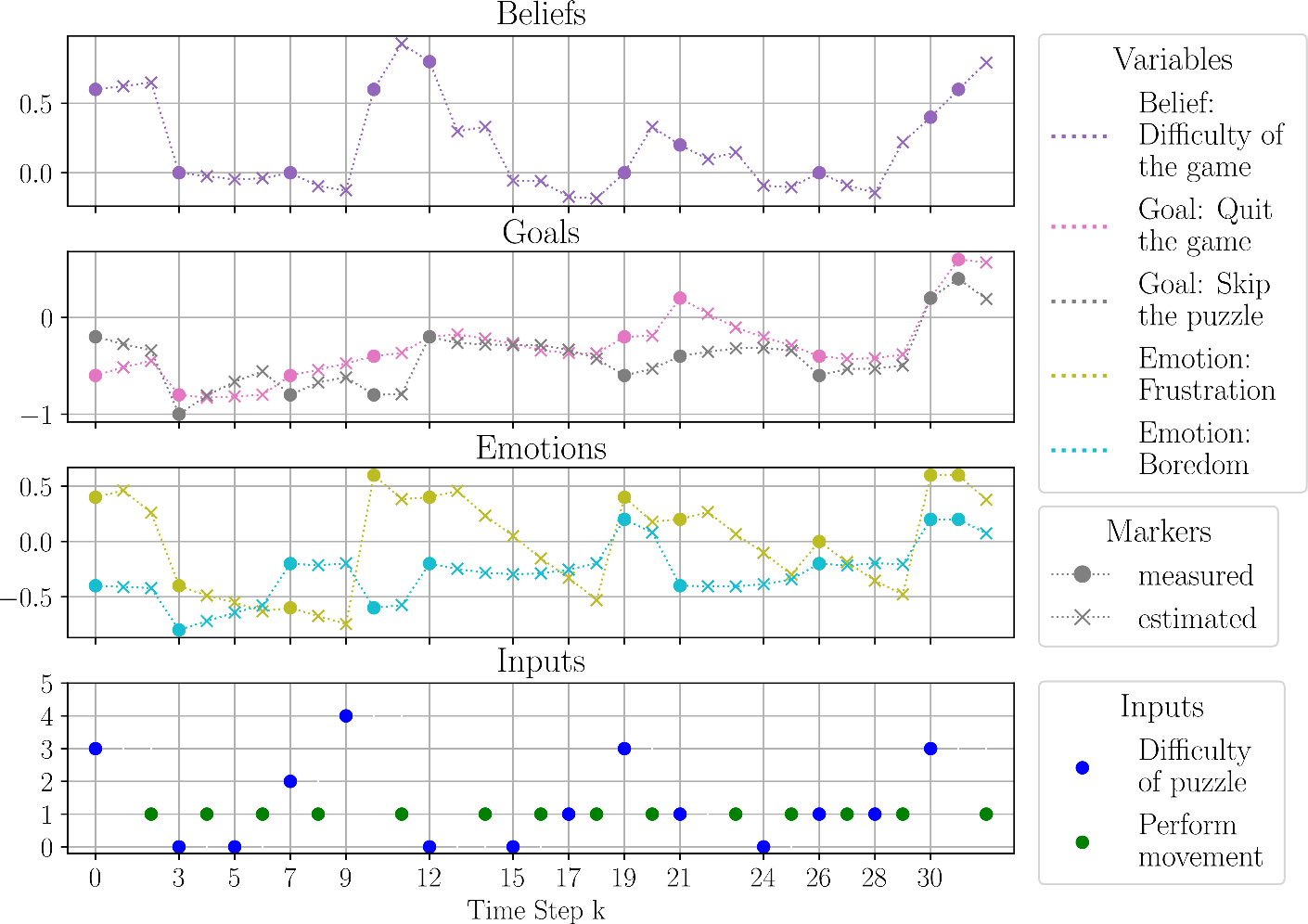}
    \caption{Values corresponding to the mental states and inputs (i.e., decision variables of the controller) for one participant over time. 
    The vertical lines correspond to the start of a new puzzle.
    Right after the start of the interaction, the participant felt frustrated, as illustrated in the plot for emotions. Hence, the controller immediately decreased the difficulty level of the next puzzle, which resulted in a decrease in the frustration of the participant. At time step $k=7$, the participant started becoming bored. Thus, for the next time step $k=9$, the controller has increased the level of difficulty of the puzzle. Again, the participant became frustrated, and the controller decreased the difficulty level of the puzzle accordingly. The controller shows to be able to sustain the engagement of the participant stably throughout the interactions by properly reacting to their mental states. 
    As for the reward, the controller has inferred that the participant prefers to always receive rewarding movements from Nao.}
    \label{fig:individual-case-2}
\end{figure*}

To showcase the performance of the proposed controller, which embeds MMM for one-step predictive decision-making (see Section~\ref{sec:MBC_controller} for details), we display in \autoref{fig:individual-case} and \autoref{fig:individual-case-2} the evolution of the belief, goals, and emotions of two participants over time, as well as the corresponding decisions (inputs) made by the controller. %
For this purpose, we used the simplified model (see \autoref{fig:cognitive_module_simple}), since this model provided better results for each participant, as described in Section~\ref{sec:case-study-model-id}. 

To evaluate the \glspl{hsri} when our control framework is used, both objective and subjective metrics were deployed. 
As explained in Section~\ref{sec:case-study-setup}, in the third session, the participants interacted with two versions of Nao — once steered by a controller based on our proposed framework and once steered by a rule-based controller that selected the inputs based on a uniform distribution. 
During each interaction in this session, we tracked the following metrics: the values for the self-reported goals and emotions of the participants, the results from self-reported questionnaires (see \autoref{fig:questionnaire} for the questionnaire that was given to the participants in the third session), and the engagement level of participants estimated from a recording of the interaction.%

\begin{figure*}[]
    \centering
    \includegraphics[width=\textwidth]{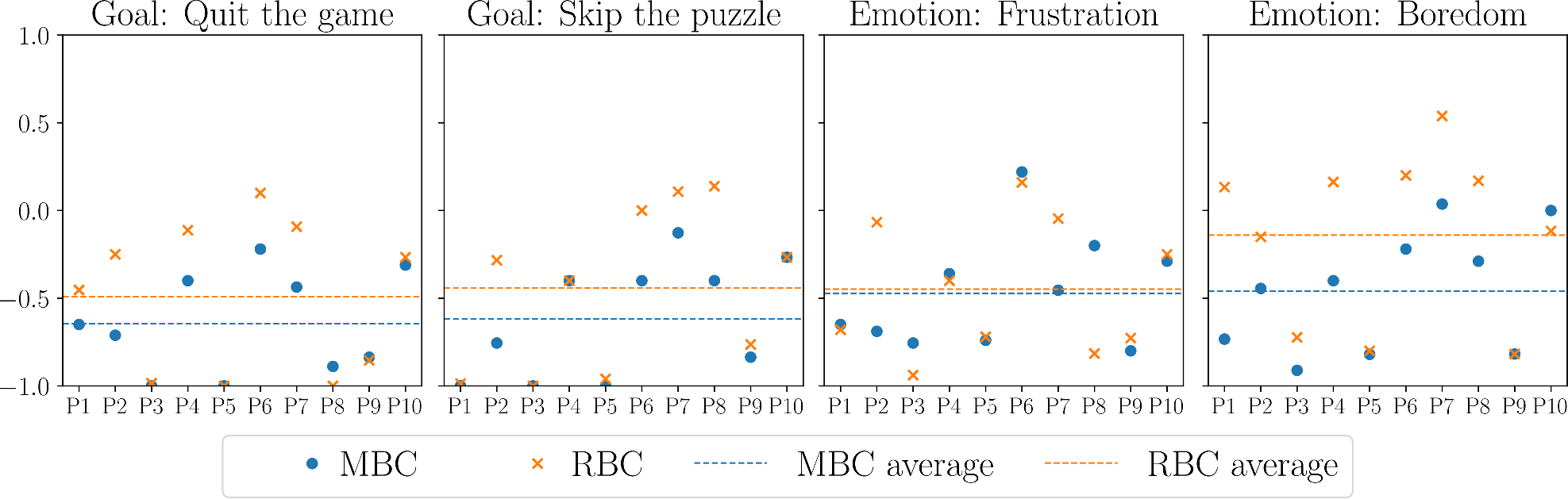}
    \caption{Average self-reported values of the emotions and goals of each participant over the interactions with the \gls{mbc} and the \alternativeController. Each graph corresponds to one variable. For a participant, the average value of that variable over the interaction with each controller is represented by a marker, whereas the dotted line represents the average value over all participants for that variable and controller (RBC stands for \alternativeController).}
    \label{fig:plots_goals_and_emotions}
\end{figure*}

The answers given by the participants regarding their emotions of ``boredom'' and ``frustration'', 
as well as their goals to ``quit the game'' and ``skip the puzzle'', were collected with the following frequency: After the first move by the participant in the first puzzle, the questions appeared on the display and the participants were asked to provide their response. In the subsequent puzzles, after the first move of the participant in the puzzle, if more than $150$ seconds passed or if the participant played $3$ puzzles since the last set of questions, then the questions were posed again. The values provided by the participants were used both to reset the values of the mental states used by MMM within the \gls{mbc} and to assess the quality of the interactions. The average emotions and goals reported by each participant over the interactions with the \gls{mbc} and with the \alternativeController\ are shown in \autoref{fig:plots_goals_and_emotions}.%

For the majority of the participants, interacting with our \gls{mbc} resulted in lower boredom (thus higher engagement) and a lower desire to quit the game or skip the puzzle. In fact, $8$ out of $10$ participants reported lower levels of boredom throughout the interaction with the \gls{mbc}, and only $1$ participant reported higher levels of boredom throughout this interaction. 
Regarding the frustration level, there does not seem to be a clear tendency when interacting with the \gls{mbc}. 
Some participants showed higher frustration levels, but less boredom with the \gls{mbc} approach (see participants P3 and P8 in \autoref{fig:plots_goals_and_emotions}). 
This suggests that, for some participants, a trade-off between frustration and engagement should be considered. An appropriate increase in the frustration level may be necessary to sustain engagement (e.g., for highly competitive people) and, thus, a longer-term interaction. 

Additionally, \autoref{tab:goals-and-emotions-values} compares the average values of the goals and emotions obtained with each controller. 
On average, when interacting with the Nao controlled by the \gls{mbc} approach, the desire of 
the participants to quit the game and to skip the puzzles were, respectively, $7.70\%$ (p-value$=0.032$) and $8.86\%$ (p-value$=0.029$) less than when the \alternativeController\ was deployed. Furthermore, participants felt $15.98\%$  (p-value$=0.009$) more engaged when interacting with Nao steered by the \gls{mbc}. There were in general no indications that using the \gls{mbc} approach was able to reduce the frustration of the participants (p-value$=0.827$).%

\begin{table}[]
\caption{Comparison between the average values of the goals and emotions for all participants when interacting with our \gls{mbc} and with the \alternativeController. The improvement of the average values per variable when the participants interacted with the \gls{mbc}, as well as the p-values, are also specified.}
\label{tab:goals-and-emotions-values}
\begin{tabular}{@{}lllll@{}}
\toprule
\multicolumn{1}{c}{\multirow{2}{*}{\textbf{Variable}}}      & \multicolumn{2}{c}{\textbf{Average}} & \multicolumn{1}{c}{\textbf{Diff.}} & \multicolumn{1}{c}{\multirow{2}{*}{\textbf{p-value}}} \\
    & \textbf{MBC} & \textbf{RBC} & \textbf{(\%)} & \\
\midrule
    \begin{tabular}[c]{@{}l@{}}$\goal{1} (k)$: Quit the game\\ \end{tabular}           & -0.645        & -0.491        & 7.70      & 0.032     \\
    \begin{tabular}[c]{@{}l@{}}$\goal{2} (k)$: Skip the puzzle\\ \end{tabular}        & -0.619        & -0.441        & 8.86      & 0.029     \\
    $\emotion{2} (k)$: Frustration                                                       & -0.472        & -0.448        & 1.17      & 0.827     \\
    $\emotion{1} (k)$: Boredom                                                           & -0.460         & -0.140         & 15.98     & 0.009     \\ \bottomrule
\end{tabular}
\end{table}

After interacting with Nao using each controller, the participant was given a questionnaire 
to assess the \gls{hsri} (see \autoref{fig:experiment_diagram-session3} for the structure of the session). The questionnaire, which can be found in \autoref{fig:questionnaire-abs}, consisted of five questions regarding the engagement, adaptability, personalization, and awareness of the robot regarding the mental states of the participant. Two questions concerned the engagement, asking participants how bored and how engaged they felt in the course of the interactions. 
The participants answered these questions using a 7-point Likert scale. 
During the debrief, preliminary participants ($2$ participants whose input was used to improve the setup of the 
experiments, but whose data was not included in the results) reported perceiving a (significant) difference in some of these categories between the two interactions deploying the \gls{mbc} and the \alternativeController. 
However, their absolute answers to the questionnaires did not reflect this difference.
When questioned about this, the preliminary participants reported that they were unsure whether they considered the same scale to answer both questionnaires due to the time gap between the sessions. They reported that, when answering the second questionnaire, they did not remember in detail the answers given in the first interaction, and that, during the second interaction, they experienced a change of perception of the assessed aspects (i.e., engagement, adaptability, etc.), which could have led them to use a different scale. 
To prevent this undesirable effect, after answering the questionnaire per session (once after each interaction), the participants were given an additional questionnaire where they could provide their responses in a relative sense, comparing 
their experiences about the two sessions. 
This comparative questionnaire is given in \autoref{fig:questionnaire-comp}. \autoref{tab:questinonaire} shows the results from the questionnaires.%

Although the average answers regarding the mental states favored the proposed \gls{mbc}, the answers given by the participants in the absolute questionnaires regarding the engagement, awareness shown by the robot about their mental states, personalization, and boredom are not conclusive (p-value$>0.05$).  Only the answers regarding the personalization of the robot demonstrate that participants experienced our proposed approach as tailored to themselves (p-value$=0.04$). Regarding the other four aspects, the comparative answers must be analyzed: 
The results indicate that $7$ participants reported feeling more engaged and less bored when interacting with the \gls{mbc}, while $3$ participants reported the opposite. 
Furthermore, $8$ participants perceived that the robot controlled by the \gls{mbc} approach was more aware of their mental states, $1$ participant had no opinion, and the other participant perceived the \alternativeController\ to display more awareness. Finally, $6$ participants perceived our \gls{mbc} approach to be more adapted to them, whereas $3$ participants stated having no opinion.%

\begin{table*}[h]
\caption{Results obtained from the two questionnaires given to participants in the third session. 
The absolute answers refer to the two questionnaires posed at the end of each interaction of the third session, while the relative answers correspond to the final questionnaire given at the end of the third session (see \autoref{fig:experiment_diagram-session3}). Note that the absolute answers were given in the range $[1,7]$. Thus, the closer the value of the positive indicators (i.e., engagement, awareness, adaptability, and personalization) are to $7$, the better. 
The value of the boredom should be as low as possible. 
In the comparative questionnaire, participants could choose between one of the two controllers or neither for each question.
For each criteria, the number of participants who have voted for each of controller (or neither) is counted. Therefore, the numbers per row add up to the total number of participants, i.e., $10$.}
\label{tab:questinonaire}
\begin{tabular*}{\textwidth}{@{\extracolsep\fill}lcccccc}
\toprule%
{\multirow{2}{*}{\textbf{Question}}}    & \multicolumn{3}{c}{Absolute Answers} & \multicolumn{3}{c}{Comparative Answers}    \\ \cmidrule{2-4}\cmidrule{5-7}%
                                        & MBC       & RBC       & p-value      & MBC & RBC & \multicolumn{1}{c}{No opinion} \\
\midrule
Q1: Engagement              & 4.7       & 4.4       & 0.496        & 7   & 3   & 0                              \\
Q2: Awareness               & 4.0       & 2.9       & 0.057        & 8   & 1   & 1                              \\
Q3: Adaptability            & 3.6       & 2.5       & 0.040         & 5   & 1   & 4                              \\
Q4: Personalization         & 3.4       & 3.0       & 0.479        & 6   & 1   & 3                              \\
Q5: Boredom                 & 3.5       & 4.9       & 0.158        & 2   & 7   & 1                              \\
\bottomrule
\end{tabular*}
\end{table*}

Finally, a video stream was used to capture the engagement of the users, using the tool presented in \cite{engagement}. 
This tool quantified the level of engagement between $0$ and $1$, where a value close to $1$ indicates that the participant is engaged.  
Although this tool correctly measured the engagement from the perspective of the robot in \cite{engagement}, the dataset where it was trained included only robot-centric interactions, i.e., the robot was the focus point. In our case, the main objective was for the robot to keep the participants engaged in solving the puzzles. Thus, we placed the camera right above the screen where the puzzles were displayed to record the participants, rather than on the robot. 
Hence, the tool indicated a high level of engagement when the participant gazed at the screen. Nevertheless, it indicated a low score regardless of whether the participant looked away from the interaction or at the robot.
The average engagement of each participant, according to this tool, is presented in \autoref{tab:engagement}. 
The results show non-significant differences between the two interactions of each participant (p-value$=0.106$) and are not coherent with the self-reported answers to the questions asked during the interactions or the questionnaire.
This discrepancy is likely due to the engagement measurement tool not being suitable for our application since it did not account for a second focus point.
As a result, this metric was not representative of the real engagement of the user in the current \gls{hsri}. 

\begin{table*}[]
\caption{Results obtained for the participants using the engagement measurement tool provided in \cite{engagement}.}
\label{tab:engagement}
\centering
\begin{tabular}{@{}llllllllllll@{}}
\toprule
Controller & P1    & P2    & P3    & P4    & P5    & P6    & P7    & P8    & P9    & P10   & Average \\ \midrule
MBC        & 0.948 & 0.962 & 0.903 & 0.859 & 0.945 & 0.841 & 0.968 & 0.884 & 0.931 & 0.923 & 0.916   \\
RBC        & 0.948 & 0.942 & 0.930 & 0.880 & 0.951 & 0.864 & 0.970 & 0.919 & 0.924 & 0.932 & 0.926   \\ \bottomrule
\end{tabular}
\end{table*}

\section{Conclusions and topics for future research}
\label{sec:conclusion}
\glsresetall

This paper introduces a novel systems-and-control theoretic framework for designing and sustaining 
\glspl{hsri}, 
ensuring desired subjective outcomes (e.g., long-term human engagement) and enriched human experiences (e.g., feeling understood by the robot), while enabling systematic adaptability driven by the robot in the interactions.%

Leveraging the recently introduced mathematical model of human perception, cognition, and decision-making, MMM \cite{paper1}, grounded in the principles of \gls{tom}, we have adapted MMM for \glspl{sr} to exhibit \gls{tom}-like behavior. 
This model has been embedded into a controller, enabling \glspl{sr} to track and adapt their interactive behavior based on the evolving mental states of their human counterpart. 
This critical capability addresses a key limitation in current \glspl{sr}, empowering them to sustain meaningful social interactions with humans \cite{Leite2013, Mataric2016} 
and to engage them for extended periods. This is expected to result in significant societal impacts, particularly for vulnerable populations \cite{Luperto2024, Khosla2019, HunLee2023, Tapus2008, Clabaough2019, Scassellati2018ImprovingRobot}.

We carried out a case study with $10$ volunteer participants 
who solved chess puzzles on a screen while interacting and being guided by a Nao robot. The case-study consisted of $3$ sessions lasting $45$ to $90$ minutes. During these sessions, the robot 
dynamically adjusted the difficulty 
level of the puzzles and rewarded participants with entertaining movements. 
In the first two sessions, data on mental states of the participants was collected while they played the chess puzzles and interacted with Nao. 
This data was then used to personalize the parameters of MMM for each participant. 
In the final session, participants interacted with Nao under two conditions: once, when the 
robot was steered by a model-based controller 
embedding their personalized MMM, and once when Nao was guided by a conventional rule-based controller 
that did not systematically consider their mental states.%

MMM achieved an average mean squared error of $0.067$ 
in tracking the beliefs, goals, and emotions of the participants within a normalized range of $[-1,1]$.
Compared to the rule-based controller, the MMM-based controller increased participant engagement 
by $16\%$ ($p=0.009$) and decreased the goal of quitting the game by $8\%$ ($p=0.032$), as 
measured objectively in the final sessions. 
Responses to a post-interaction questionnaire further confirmed that most participants perceived  Nao, when controlled by the MMM-based controller, to be more engaging, more aware of and adaptive to their mental states, and more personalized to their needs.%

Although the MMM-based controller successfully adjusted the difficulty level of the puzzles 
based on the mental states of the participants in the final session, it did not do the same for rewarding them. 
One possible reason is that during the first two sessions, while the novelty effect 
(i.e., a temporary increase in the engagement and interest of individuals when exposed to 
new experiences, e.g., interacting with a Nao robot) was still present, 
participants consistently reacted positively to the entertaining movements of Nao. 
This may have led MMM to infer that these movements always had either a positive or a negative influence on the participants, regardless of their actual mental states. 
Furthermore, while MMM demonstrated excellent performance in short-term predictions of mental states, 
its long-term estimations 
declined for some participants.

Future work should explore the impact of the novelty effect 
on collected data and enhance the state estimation of MMM to reduce 
reliance on direct measurements. Furthermore, the approach to measure the engagement of the participants should include one camera per focus point, and merge the results from all cameras,  to gather an accurate estimate of the engagement of the user. Finally, a personality test could be administered to the participants prior to the interaction to establish which mental states are relevant to optimize (e.g., whether the frustration might boost or hinder the engagement of that participant). 
Overall, this paper is the first to explore systems-and-control-theoretic methods for \glspl{hsri}, providing guarantees on performance and safety. It highlights the potential of control theory and model-based approaches in steering the behavior of \glspl{sr}.


\backmatter

\bmhead{Acknowledgments}
This research has been supported by the TU Delft AI Labs \& Talent programme.

\bmhead{Data availability}
The data generated and analyzed in this study is available in an anonymized format in \url{https://doi.org/10.4121/ccadc914-9502-46d6-9ba5-fef581f2933f}. This includes the anonymized raw data collected during experiments and the processed data derived from it. However, video recordings of participants are not publicly available due to restrictions outlined in the informed consent forms.

\bmhead{Code availability}
The code used to generate the results in this study is available at \url{https://github.com/marialuis-mp/MMM-Controller-for-Social-Robot}.

\section*{Declarations}

\bmhead{Ethical approval declarations}
All experiments involving human participants described in this article were approved by the Human Research Ethics Committee of TU Delft under \textit{Approval No. 3780}. 
The procedures of data collection were GDPR compliant and an informed consent was obtained from all participants. 

\bmhead{Conflict of interest}
The authors declare that they have no competing
financial interests or personal relationships that could have appeared to
influence the work reported in this paper
\noindent

\begin{appendices}

\onecolumn
\newpage

\section{Questions posed to the participants during the interactions with Nao}
\label{app:questions}

\begin{figure}[h!]
     \centering
     \begin{subfigure}[h]{0.49\textwidth}
         \centering
         \includegraphics[width=\textwidth]{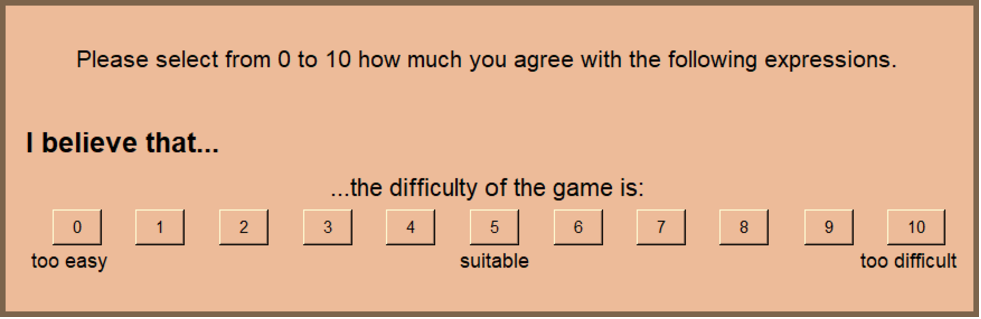}
         \caption{Beliefs}
         \label{fig:qb}
     \end{subfigure}
      \hfill
     \begin{subfigure}[h]{0.49\textwidth}
         \centering
         \includegraphics[width=\textwidth]{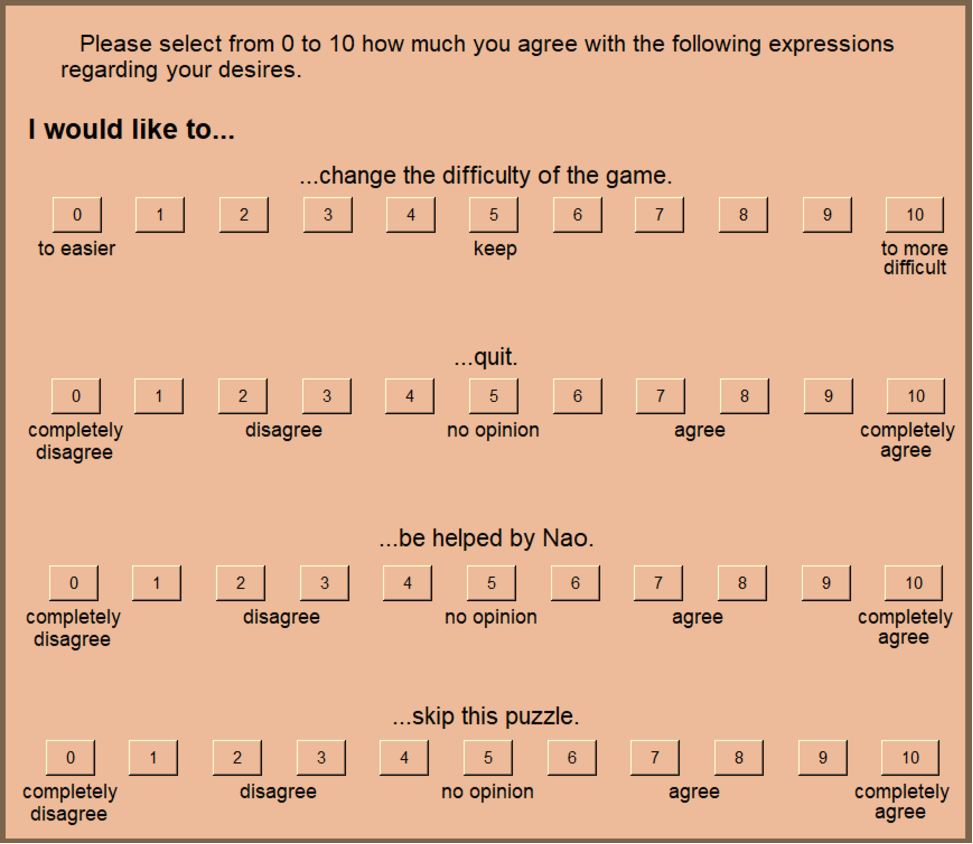}
         \caption{Goals}
         \label{fig:qg}
     \end{subfigure}
     \hfill
     \begin{subfigure}[h]{0.49\textwidth}
         \centering
         \includegraphics[width=\textwidth]{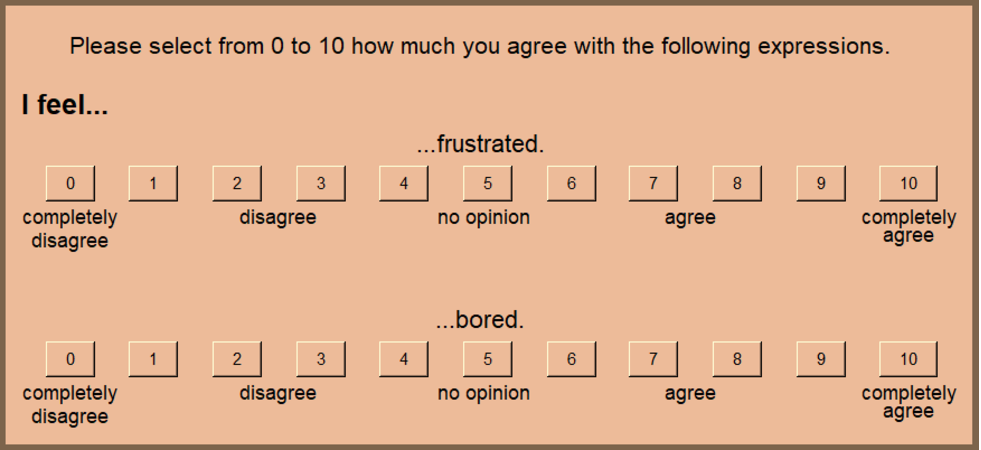}
         \caption{Emotions}
         \label{fig:qe}
     \end{subfigure}
      \hfill
     \begin{subfigure}[h]{0.49\textwidth}
         \centering
         \includegraphics[width=\textwidth]{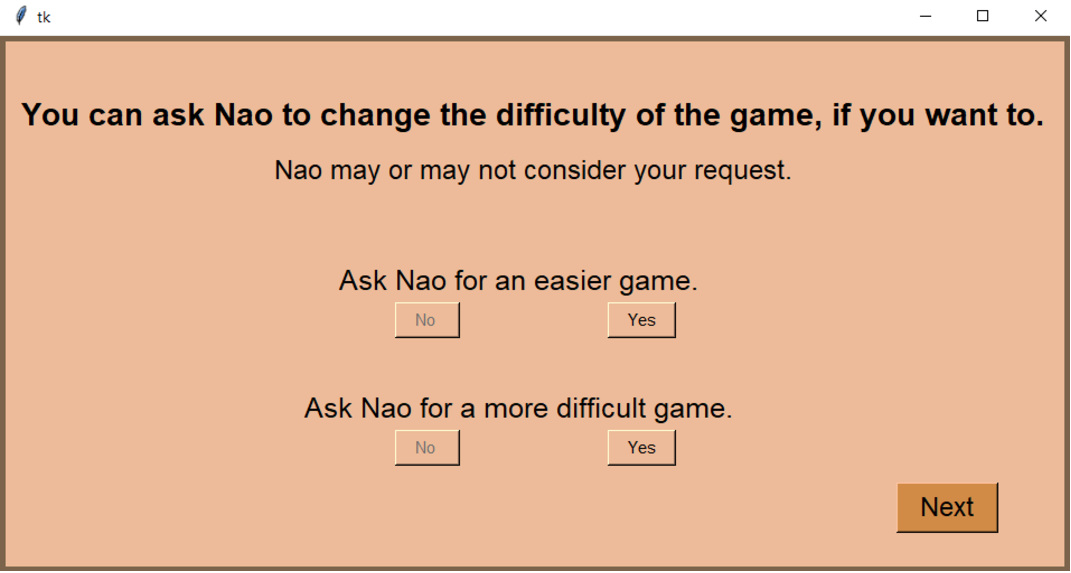}
         \caption{Actions}
         \label{fig:qa}
     \end{subfigure}
        \caption{Periodic questions asked to the participants during the interaction with the Nao robot. In the first and second sessions, they enabled the collection of the data used to identify the MMM per participant. In the last session, they were used to assess the interaction and to update the state variables of the MMM that was integrated in the controller of Nao.}
        \label{fig:questions}
\end{figure}

This appendix includes the questions asked to the participants during their interactions with the Nao robot. The questions were posed to the participants to gather information about their beliefs (\autoref{fig:qb}), goals (\autoref{fig:qg}), and emotions (\autoref{fig:qe}). 
Furthermore, at the end of each puzzle, participants were presented with the questions shown in \autoref{fig:qa}, allowing them to request an easier or more difficult puzzle. However, while participants could make this request, Nao did not fulfill it. Instead, the request was used to gather information about the participant's actions.

\vspace{1cm}
\section{Difficulty levels of the chess puzzles}\label{app:difficulty}

This appendix includes information on how the difficulty of the puzzles was selected throughout the interactions with the Nao robot. \autoref{tab:ratings-per-diff} describes the conversion between the ratings attributed to the puzzles in the  \textit{lichess.org} database \cite{lichess_db}, where the puzzles were retrieved from, and the categorization in six difficulty levels adopted by the controller of Nao.
Moreover,  \autoref{tab:diff-per-interactions} shows the difficulty level of each puzzle proposed to the participants in the first two interactions. 
\begin{table}[h!]
\caption{Difficulty levels of the chess puzzles and the corresponding minimum and maximum ratings attributed in the \textit{lichess.org} database \cite{lichess_db}.}
\label{tab:ratings-per-diff}
\begin{tabular}{@{}lll@{}}
\toprule
Puzzle difficulty level & Minimum Rating & Maximum Rating \\ 
\midrule
0 & 600 & 800 \\
1 & 920 & 1120 \\
2 & 1240 & 1440 \\
3 & 1560 & 1760 \\
4 & 1880 & 2080 \\
5 & 2200 & 2400 \\
\bottomrule
\end{tabular}
\end{table}
\begin{table}[h]
\centering
\caption{Difficulty levels of the chess puzzles in the first two experimental sessions with the participants.}
\label{tab:diff-per-interactions}   
\begin{tabular}{@{}lll@{}}
\toprule
Order number of the puzzle & Difficulty level in & Difficulty level in \\
offered to the participant & session 1 & session 2 \\ 
\midrule
1, 2, 3 & 0 & 5 \\
4, 5, 6 & 2 & 3 \\
7, 8, 9 & 4 & 1 \\
10, 11, 12 & 4 & 1 \\
13, 14, 15 & 2 & 3 \\
16, 17, 18\footnotemark[1]  & 0 & 5 \\ \hline 
19, 20, 21 & 0 (as in the $1^{\text{st}}$ row) & 5 (as in the $1^{\text{st}}$ row)  \\
21, 23, 24 & 2 (as in the $2^{\text{nd}}$ row) & 3 (as in the $2^{\text{nd}}$ row)  \\
... & ... & ... \\
\bottomrule
\end{tabular}
\footnotetext[1]{End of the minimum recommended number of puzzles for each session.}
\end{table}

\section{Reward movements performed by Nao in the case study}\label{app:reward-movements}
\begin{figure}[h!]
    \centering
    \vspace{0.2cm}
    \begin{subfigure}[h]{\textwidth}
        \includegraphics[width=\linewidth]{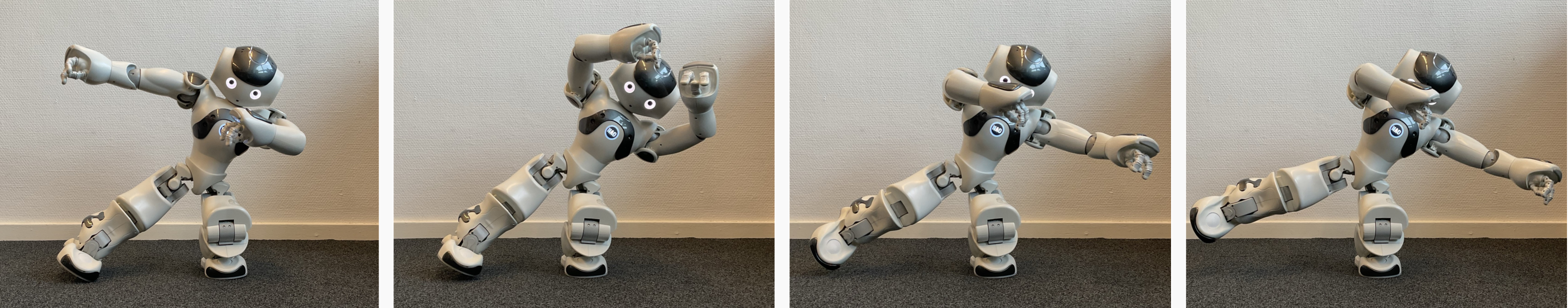}
        \caption{Performing Tai Chi}
        \label{fig:movement-tai-chi}
    \end{subfigure}
    \hfill
    \begin{subfigure}[h]{0.66\textwidth}
        \centering
        \includegraphics[width=\linewidth]{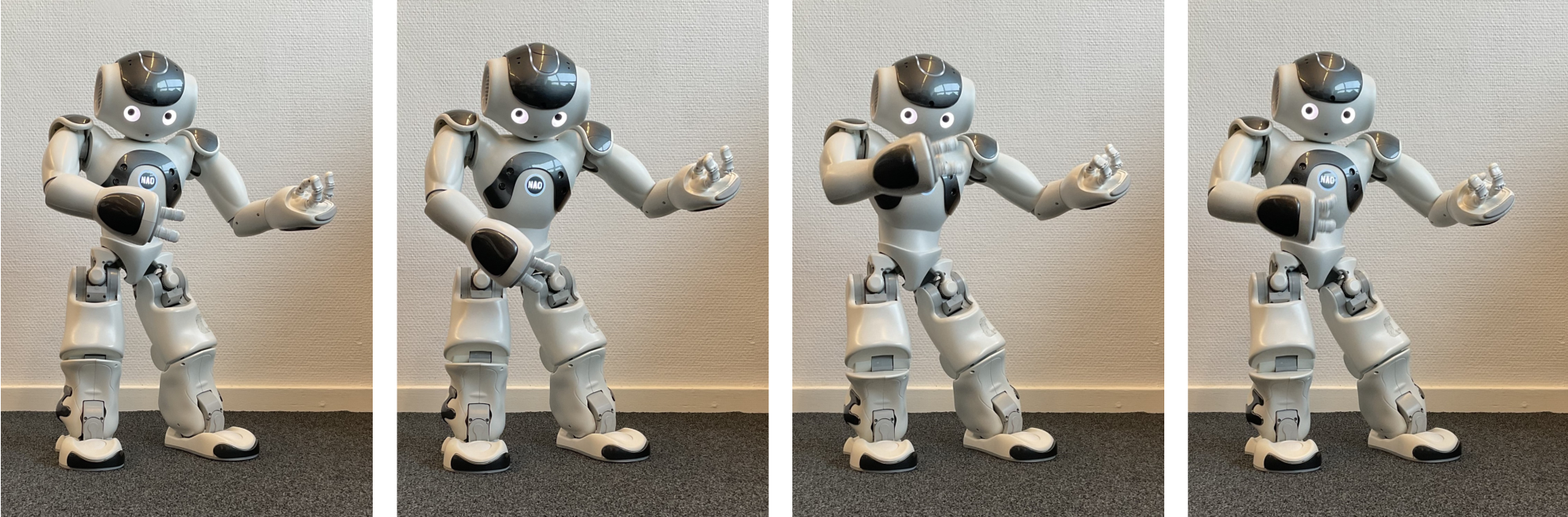}
        \caption{Playing the guitar}
        \label{fig:movement-guitar}
    \end{subfigure}
    \hfill
    \begin{subfigure}[h]{0.29\textwidth}
        \centering
        \includegraphics[width=\linewidth]{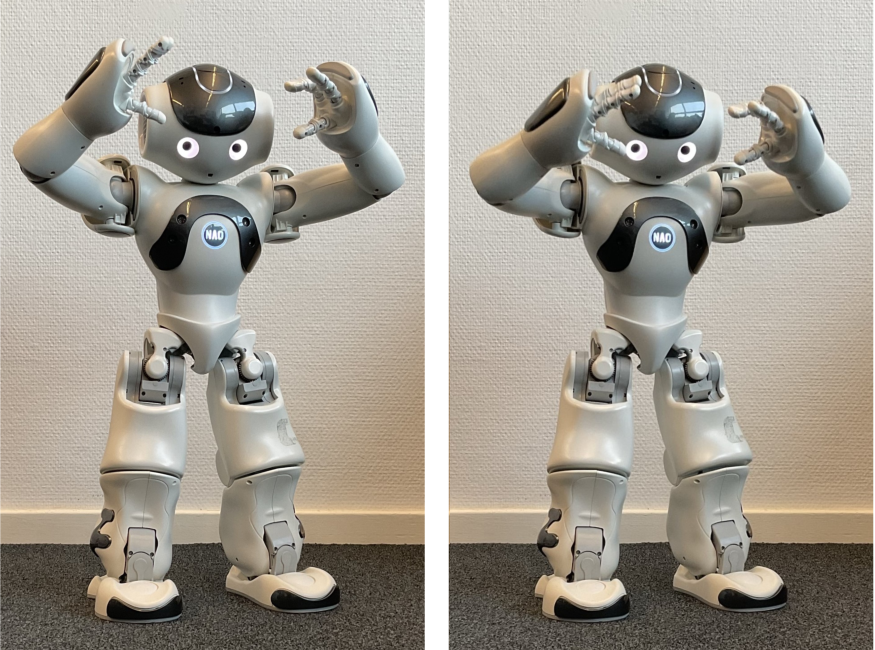}
        \caption{Taking a photo}
        \label{fig:movement-photo}
    \end{subfigure}
    \hfill
    \begin{subfigure}[h]{0.54\textwidth}
        \centering
        \includegraphics[width=\linewidth]{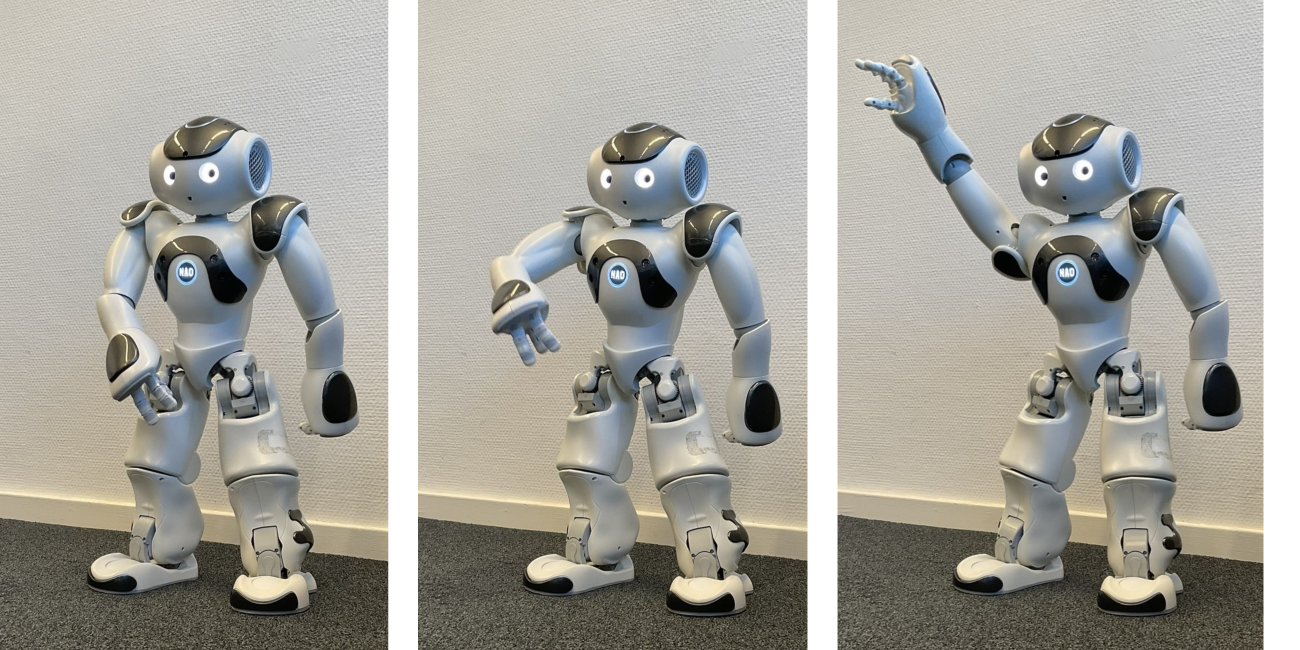}
        \caption{Dancing}
        \label{fig:movement-dance}
    \end{subfigure}
    \hfill
        \begin{subfigure}[h]{0.43\textwidth}
        \centering
        \includegraphics[width=\linewidth]{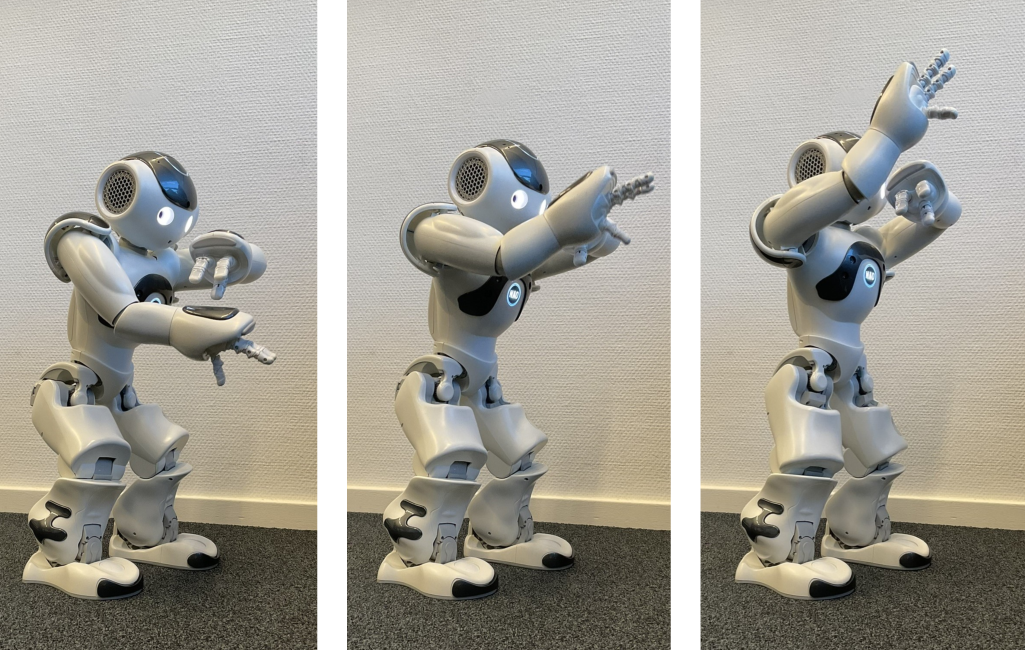}
        \caption{Mimicking an elephant}
        \label{fig:movement-elephant}
    \end{subfigure}
    \caption{The five reward movements performed by the Nao robot in the experimental sessions with the participants.}
        \label{fig:movements}
\end{figure}

This appendix contains a visual representation of the five entertaining movements performed by the Nao during experimental sessions of the case study. 
These movements were performed by the Nao robot to reward the participants and keep them engagement in the interaction. 
These five movements included performing tai chi (\autoref{fig:movement-tai-chi}), playing the guitar (\autoref{fig:movement-guitar}), pretending to take a photo (\autoref{fig:movement-photo}), dancing (\autoref{fig:movement-dance}), and pretending to be an elephant (\autoref{fig:movement-elephant}).


\section{Structure of the modules of the MMM used in the case study}\label{app:model}

\begin{figure}[h!]
    \centering
    \includegraphics[width=0.85\textwidth]{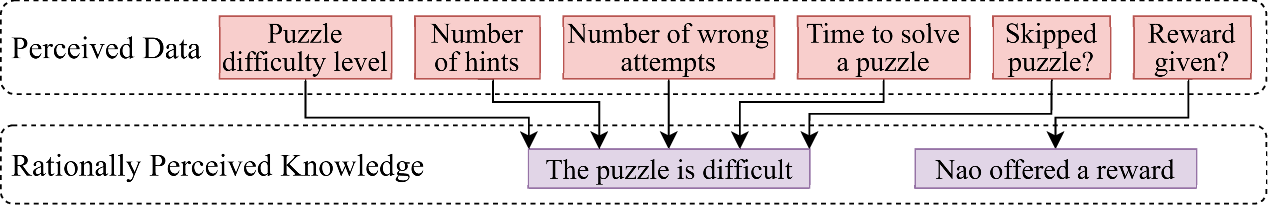}
    \caption{Rational reasoning sub-process of the perception module of the \gls{tom} model (MMM) used in the case study.}
    \label{fig:perception_module} 
\end{figure}

\begin{figure}[h!]
    \centering
    \includegraphics[width=0.8\textwidth]{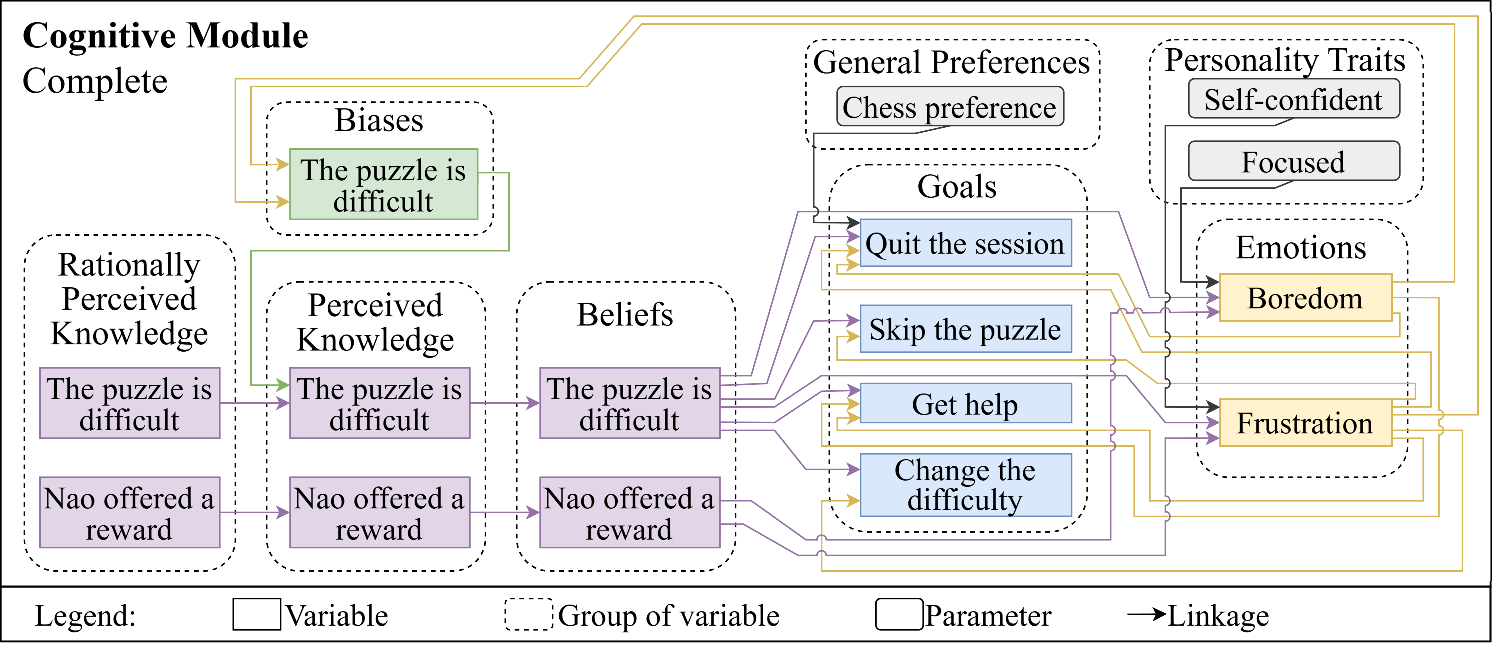}
    \caption{Variables and linkages of the cognitive module of the \gls{tom} model (MMM) used in the case study.}
    \label{fig:cognitive_module_complete} 
\end{figure}

\begin{figure}[h!]
    \centering
    \includegraphics[width=0.8\textwidth]{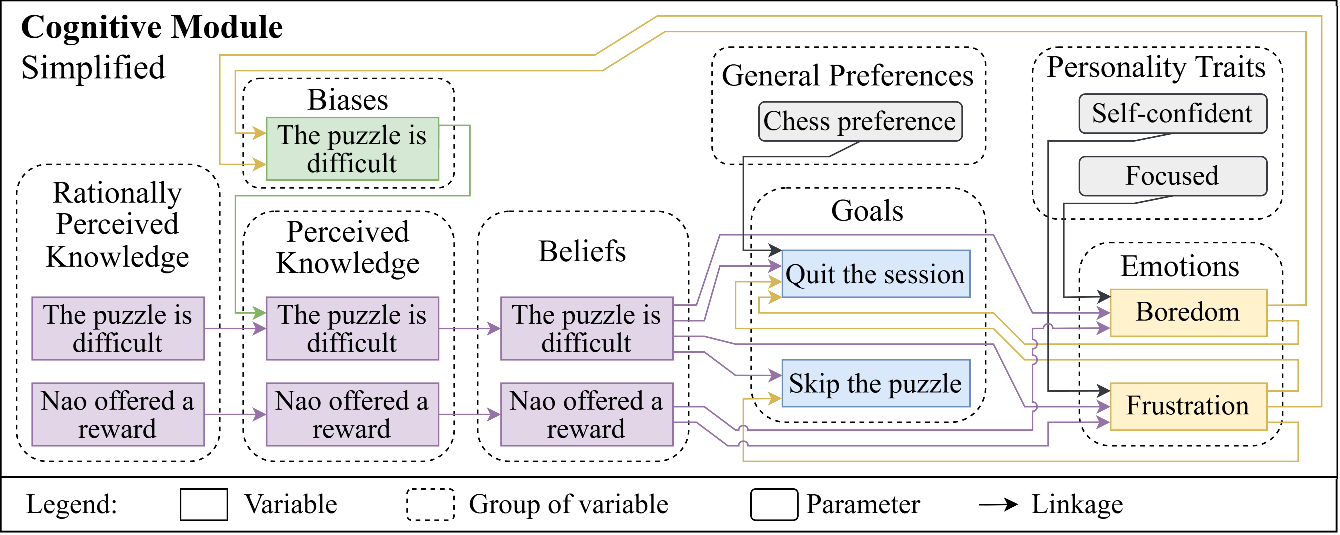}
    \caption{Variables and linkages of the simplified cognitive module of the \gls{tom} model (MMM) used in the case study.}
    \label{fig:cognitive_module_simple} 
\end{figure}

This appendix includes graphical representations of part of the perception module (\autoref{fig:perception_module}), the cognition module (\autoref{fig:cognitive_module_complete} and \autoref{fig:cognitive_module_simple}), and part of the \dm\ module (see \autoref{fig:decision-making}). 
These figures showcase the variables that are part of each module, as well as which variables influence each other (these influences are represented by arrows). 

Furthermore, two structures are displayed for the cognitive module, \autoref{fig:cognitive_module_complete} and \autoref{fig:cognitive_module_simple}. While \autoref{fig:cognitive_module_complete} represents the model used in the first two sessions with the participants, \autoref{fig:cognitive_module_simple} shows the model used in the last session for the integration within the model-based controller. 
For clarity, the influencers of the variables in the cognition module are also given in \autoref{tab:cognition-module-influencers}.

\begin{figure}[h!]
    \centering
    \includegraphics[width=\textwidth]{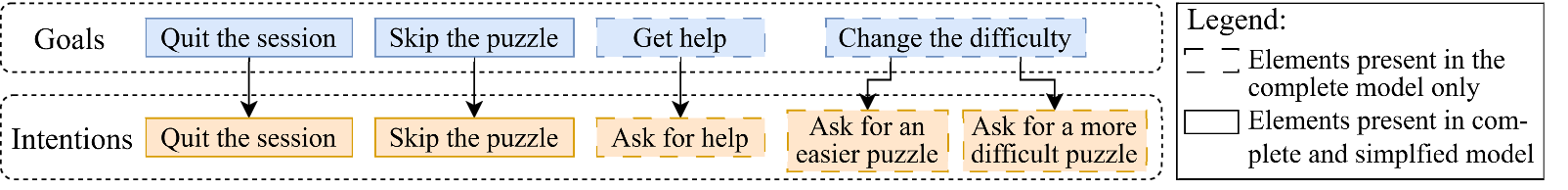}
    \caption{Variables and linkages of the rational intention selection of the \dm\space module of the \gls{tom} model (MMM) used in the case study.}
    \label{fig:dm_module} 
\end{figure}

\begin{table}[h!]
    \centering
    \caption{The influencer variables of each one of the variables of the cognition module.}
    \label{tab:cognition-module-influencers}
    \begin{tabular}{ll}
         \toprule
         Variable and concept & Influencers (and concepts)  \\
         \midrule
         $\belief{1} (k)$ (The puzzle is difficult) & $\perceivedKnowledge{1} (k)$ (The puzzle is difficult) \\
         $\belief{2} (k)$ (Nao offered a reward)    & $\perceivedKnowledge{2} (k)$ (Nao offered a reward)    \\
         $\goal{1} (k)$ (Quit the session)          & $\belief{1} (k)$ (The puzzle is difficult), $\emotion{1} (k)$ (Boredom), and $\emotion{2} (k)$ (Frustration) \\
         $\goal{2} (k)$ (Skip the puzzle)           & $\belief{1} (k)$ (The puzzle is difficult) and $\emotion{2} (k)$ (Frustration) \\
         $\goal{3} (k)$ (Get help)                  & $\belief{1} (k)$ (The puzzle is difficult), $\emotion{1} (k)$ (Boredom), and $\emotion{2} (k)$ (Frustration) \\
         $\goal{4} (k)$ (Change difficulty)         & $\belief{1} (k)$ (The puzzle is difficult) and $\emotion{2} (k)$ (Frustration)\\
         $\emotion{1} (k)$ (Boredom)                & $\belief{1} (k)$ (The puzzle is difficult) and $\belief{2} (k)$ (Nao offered a reward) \\
         $\emotion{2} (k)$ (Frustration)            & $\belief{1} (k)$ (The puzzle is difficult) and $\belief{2} (k)$ (Nao offered a reward) \\
         $\bias (k)$ (The puzzle is difficult) & $\emotion{1} (k)$ (Boredom) and $\emotion{2} (k)$ (Frustration) \\
         $\perceivedKnowledge{1} (k)$ (The puzzle is difficult) & $\rationallyPerceivedKnowledge{1} (k)$ (The puzzle is difficult) and $\bias (k)$ (The puzzle is difficult)\\
         $\perceivedKnowledge{2} (k)$ (Nao offered a reward)    & $\rationallyPerceivedKnowledge{2} (k)$ (Nao offered a reward)    \\
         \bottomrule
    \end{tabular}
\end{table}

\section{Questionnaire}\label{app:questionnaire}

This appendix provides the questionnaires that were given to the participants in session~3. 

\begin{figure}[h!]
     \centering
     \begin{subfigure}[h]{0.49\textwidth}
         \centering
         \includegraphics[width=\textwidth]{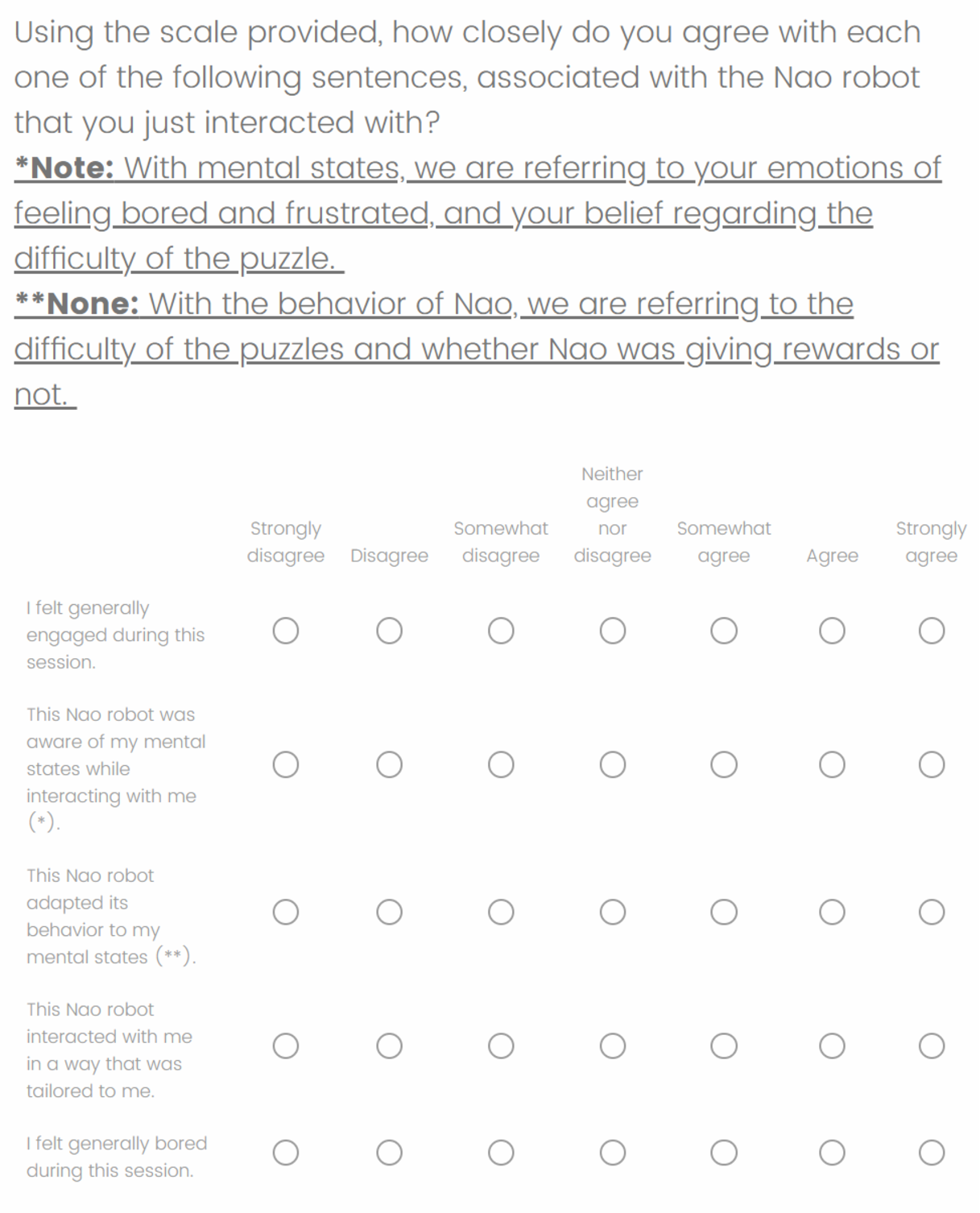}
         \caption{Absolute questionnaire given to the participants after each one of the two interactions of the third session.}
         \label{fig:questionnaire-abs}
     \end{subfigure}
      \hfill
     \begin{subfigure}[h]{0.49\textwidth}
         \centering
         \includegraphics[width=\textwidth] {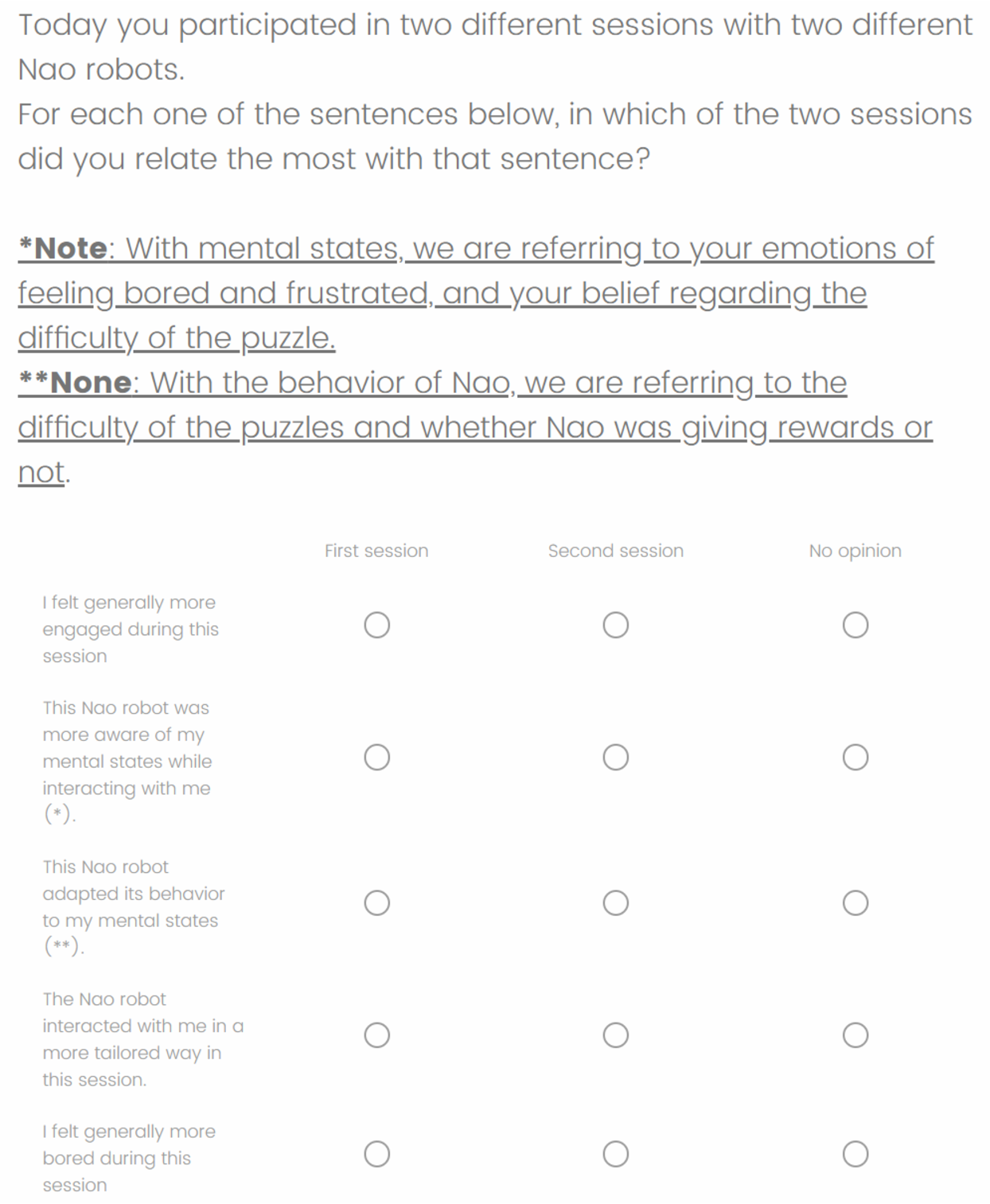}
         \caption{Comparative questionnaire given to the participants at the end of the third session, comparing the two interactions.}
         \label{fig:questionnaire-comp}
     \end{subfigure}
        \caption{Questionnaires given to the participants during the third session.}
        \label{fig:questionnaire}
\end{figure}




\end{appendices}

\newpage

\bibliography{sn-bibliography}

\end{document}